\documentclass[twocolumn,prb,groupedaddress,showpacs,preprintnumbers,amsmath,amssymb,floatfix]{revtex4}
\usepackage{graphicx}

\begin{document}

\title{Why do ultrasoft repulsive particles cluster and crystallize?\\
Analytical results from density functional theory}

\author{Christos N.\ Likos}
\email{likos@thphy.uni-duesseldorf.de}
\affiliation{Institut f\"ur Theoretische Physik II: Weiche Materie,
Heinrich-Heine-Universit\"at D{\"u}sseldorf,
Universit{\"a}tsstra{\ss}e 1, D-40225 D\"{u}sseldorf,
Germany}

\author{Bianca M.\ Mladek}
\affiliation{Center for Computational Materials Science and Institut f\"ur
  Theoretische Physik, Technische Universit\"at Wien, Wiedner Hauptstra{\ss}e
  8-10, A-1040 Wien, Austria}

\author{Dieter Gottwald} 
\affiliation{Center for Computational Materials Science and Institut f\"ur
  Theoretische Physik, Technische Universit\"at Wien, Wiedner Hauptstra{\ss}e
  8-10, A-1040 Wien, Austria}

\author{Gerhard Kahl}
\affiliation{Center for Computational Materials Science and Institut f\"ur
  Theoretische Physik, Technische Universit\"at Wien, Wiedner Hauptstra{\ss}e
  8-10, A-1040 Wien, Austria}

\date{\today}

\begin{abstract}
We demonstrate the accuracy of the hypernetted chain closure and of
the mean-field approximation for the calculation of the
fluid-state properties of systems interacting by means of bounded
and positive-definite pair potentials with oscillating Fourier
transforms. Subsequently, we prove the validity of a bilinear,
random-phase density functional for arbitrary inhomogeneous phases
of the same systems. On the basis of this functional, we calculate
analytically the freezing parameters of the latter. We demonstrate
explicitly that the stable crystals feature a lattice constant that
is independent of density and 
whose value is dictated by the position of the negative minimum of
the Fourier transform of the pair potential.
This property is equivalent with the
existence of clusters, whose population scales proportionally to
the density. We establish that
regardless of the form of the interaction
potential and of the location on the freezing line, all 
cluster crystals have a universal Lindemann ratio $L_{\rm f} =
0.189$ at freezing. 
We further make an explicit link between the 
aforementioned density functional and the harmonic theory of crystals.
This allows us to establish an equivalence between the 
emergence of clusters and the existence of negative
Fourier components of the interaction potential.
Finally, we make a connection between the 
class of models at hand and the system of infinite-dimensional
hard spheres, when the limits of interaction steepness and
space dimension are both taken to infinity in a particularly
described fashion.
\end{abstract}


\maketitle

\section{Introduction}
\label{intro:sec}

Cluster formation in complex fluids is a topic that has
attracted considerable attention 
recently.\cite{sear,fs1,fs2,bartlett1,bartlett2,lr1,lr2,jpcm,epl}
The general belief is that a short-range attraction in the
pair interaction potential is necessary to initiate 
aggregation and a long-range repulsive tail in order to
limit cluster growth and prevent phase separation. 
An alternative scenario for cluster formation pertains to systems whose 
constituent particles interact by means of 
purely repulsive potentials. Cluster formation in this case
is counterintuitive at first sight: why should particles form clusters
if there is no attraction acting between them? The answer lies
in an additional property of the effective repulsion, namely
that of being {\it bounded}, thus allowing full particle overlaps.
Though surprising and seemingly unphysical at first, bounded
interactions are fully legitimate and natural as effective
potentials\cite{likos:pr:01}
between polymeric macromolecular aggregates of low internal 
monomer concentration, such as 
polymers,\cite{krueger:89, hall:94, louis:prl:00}
dendrimers,\cite{ingo:jcp:04,ballik:ac:04} 
microgels,\cite{denton:03,gottwald:04,gottwald:05}
or coarse-grained block copolymers.\cite{pierleoni:prl:06,hansen:molphys:06} 
The growing
interest in this type of effective interactions is also underlined
by the recent mathematical proof of the existence of crystalline
ground states for such systems.\cite{suto:prl:05,suto:prb:06}

Cluster formation in the fluid {\it and} in the crystal phases
was explicitly seen in the system of penetrable spheres, 
following
early simulation results\cite{klein:94} and 
subsequent cell-model calculations.\cite{pensph:98} 
Cluster formation was
attributed there to the tendency of particles to 
create free space by 
forming full overlaps. The conditions under which ultrasoft
and purely repulsive particles form clusters have been conjectured
a few years ago\cite{criterion} and explicitly confirmed
by computer simulation very recently.\cite{bianca:prl:06}
The key lies in the behavior of
the Fourier transform of the effective interaction potential:
for clusters to form, it must contain negative parts, forming
thus the class of $Q^{\pm}$-interactions. The complementary
class of potentials with purely nonnegative Fourier transforms,
$Q^{+}$, does not lead to clustering but to remelting at high
densities.\cite{stillinger:physica,likos:gauss,saija1, saija2, saija3, archer}  
An intriguing feature of the crystals
formed by $Q^{\pm}$-systems is the independence of the 
lattice constant on density,\cite{criterion,bianca:prl:06}
a feature that reflects the flexibility of soft matter
systems in achieving forms of self-organization unknown
to atomic ones.\cite{daan:nature,daan:web}
The same characteristic has recently been seen also
in slightly modified models that contain a short-range 
hard core.\cite{primoz:07} 
In this work, we provide an analytical
solution of the crystallization problem and of the properties
of the ensuing solids within the framework of an accurate
density functional approach. We explicitly demonstrate the 
persistence of a single length scale at all densities and
for all members of the $Q^{\pm}$-class and offer thus broad
physical insight into the mechanisms driving the stability
of the clustered crystals. We further establish some universal
structural properties of all $Q^{\pm}$-systems both in the
fluid and in the solid state, justifying the use of the 
mean-field density functional on which this work rests.
We make a connection between our results and the harmonic
theory of solids in the Einstein-approximation. 
Finally, we establish a connection with suitably-defined
infinite dimensional models of hard spheres.

The rest of this paper is organized as follows:
in Sec.\ \ref{dft:sec} we derive an accurate density
functional by starting with the uniform phase and establishing
the behavior of the direct correlation functions of the fluid
with density and temperature. Based on this density functional,
we perform an analytical calculation
of the freezing characteristics of the $Q^{\pm}$-systems by
employing a weak approximation in Sec.\ \ref{analytical:sec}.
The accuracy of this approximation is successfully tested against full
numerical minimization of the functional in Sec.\ \ref{compare:sec}.
In Sec.\ \ref{harmonic:sec} the equivalence between the density
functional and the theory of harmonic crystals is demonstrated,
whereas in Sec.\ \ref{invpower:sec} a connection is made with
inverse-power potentials. Finally, in Sec.\ \ref{summary:sec}
we summarize and draw our conclusions. Some intermediate, technical
results that would interrupt the flow of the text are relegated
in the Appendix. 

\section{Density functional theory}
\label{dft:sec}

\subsection{Definition of the model}
\label{model:sec}

In this work, we focus our interest on systems of spherosymmetric
particles
interacting by means of {\it bounded} 
pair interactions $v(r)$ of the form:
\begin{equation}
v(r) = \epsilon\phi(r/\sigma),
\label{gen_poten:eq}
\end{equation}
with an energy scale $\epsilon$ and a length scale $\sigma$,
and which fulfill the Ruelle conditions for stability.\cite{Ruelle:book} In
Eq.\ (\ref{gen_poten:eq}) above, $\phi(x)$ is some dimensionless function 
of a dimensionless variable and $r$ denotes the distance between
the spherosymmetric particles. In the context of soft matter physics,
$v(r)$ is an effective potential between, e.g., the centers of
mass of macromolecular entities, such as polymer chains or
dendrimers.\cite{ballik:ac:04} As the centers of mass 
of the aggregates can fully overlap without this
incurring an infinitely prohibitive cost in (free) energy, the
condition of boundedness is fulfilled:
\begin{equation}
0 \leq v(r) < K,
\label{bounded:eq}
\end{equation}
with some constant $K$.

The interaction range is set by 
$\sigma$, typically the physical size 
(e.g., the gyration radius) of the macromolecular 
aggregates that feature $v(r)$ as their effective interaction.
In addition to being bounded, the second requirement to be
fulfilled by the function $\phi(x)$ is that it decay sufficiently
fast to zero as $x \to \infty$, so that its Fourier transform 
$\tilde\phi(y)$ exists. In three spatial dimensions, we have
\begin{equation}
\tilde\phi(y) = 4\pi
\int_0^{\infty}{\rm d}x\,\frac{\sin(y x)}{yx} x^2 \phi(x)
\label{ftphi:eq}
\end{equation}
and, correspondingly,
\begin{equation}
\tilde v(k) = \epsilon\sigma^3\tilde\phi(k\sigma)
\label{ftv:eq} 
\end{equation}
for the Fourier transform $\tilde v(k)$ of the potential, evaluated
at wavenumber $k$. Our focus in this work is on systems for which
$\tilde v(k)$ is oscillatory, i.e., $v(r)$ features positive
and negative Fourier components, classifying it as a
$Q^{\pm}$-potential.\cite{criterion} Though this work is general,
for the purposes of demonstration of our results, we consider
a particular realization of $Q^{\pm}$-potentials, namely the
generalized exponential model of exponent $m$, (GEM-$m$): 
\begin{equation}
v(r) = \epsilon\exp[-(r/\sigma)^m],
\label{gemn:eq}
\end{equation}
with $m=4$. It can be shown that all members of the GEM-$m$ family
with $m > 2$ belong to the $Q^{\pm}$-class, see Appendix A.
 
A short account
of the freezing and clustering behavior of the GEM-4 model has been
recently published.\cite{bianca:prl:06}
Another prominent member of the family
is the $m=\infty$ model, which corresponds to 
penetrable spheres\cite{klein:94,pensph:98,fernaud}
with a finite overlap energy penalty 
$\epsilon$. Indeed, the explicitly
calculated phase behavior of these two show strong resemblances,
with the phase diagram of both being dominated by the phenomenon
of formation of clusters of overlapping particles 
and the subsequent ordering of the same in
periodic crystalline arrangements.\cite{pensph:98,fernaud}
In this work, we provide a generic,
accurate, and analytically tractable theory of 
inhomogeneous phases of $Q^{\pm}$-systems.

\subsection{The uniform fluid}
\label{uniform:sec}

Let us start from the simpler system of a homogeneous fluid,
consisting of $N$ spherosymmetric particles enclosed in a 
macroscopic volume $V$. 
The structure and thermodynamics of the system are determined by the
density $\rho = N/V$ and the absolute temperature $T$ or, better,
their dimensionless counterparts:
\begin{eqnarray}
\nonumber
\rho^{*} \equiv \rho\sigma^3
\end{eqnarray}
and
\begin{eqnarray}
\nonumber
T^{*} \equiv \frac{k_{\rm B}T}{\epsilon},
\end{eqnarray}
with Boltzmann's constant $k_{\rm B}$. As usual, we also introduce
for future convenience the inverse temperature $\beta = (k_{\rm B}T)^{-1}$.
We seek for appropriate and accurate closures to the Ornstein-Zernike
relation\cite{hansen:book} 
\begin{equation}
h(r) = c(r) + \rho\int{\rm d}^3r'c(|{\bf r}-{\bf r'}|)h(r'),
\label{oz:eq}
\end{equation}
connecting the total correlation function $h(r)$ to the direct
correlation function $c(r)$ of the uniform fluid. One possibility
is offered by the hypernetted chain closure (HNC) that reads 
as\cite{hansen:book}
\begin{equation}
h(r) = \exp\left[-\beta v(r) + h(r) - c(r)\right] - 1.
\label{hnc:eq}
\end{equation}
An additional closure, the mean-field approximation (MFA), 
was also employed and will be discussed later.

Our solution by means of approximate closures was accompanied
by extensive $NVT$-Monte Carlo (MC) simulations. We measured the radial
distribution function $g(r) \equiv h(r) + 1$ as well as the
structure factor $S(k) = 1 + \rho {\tilde h}(k)$, 
where ${\tilde h}(k)$ is the Fourier transform of $h(r)$, to provide an assessment
of the accuracy of the approximate theories. We typically simulated
ensembles of up to $3\,000$ particles for a total of $150\,000$
Monte Carlo steps. Measurements were taken
in every tenth step after equilibration.

\begin{figure}
      \begin{center}
      \begin{minipage}[t]{8.3cm}
      \includegraphics[width=8.2cm, clip]
        {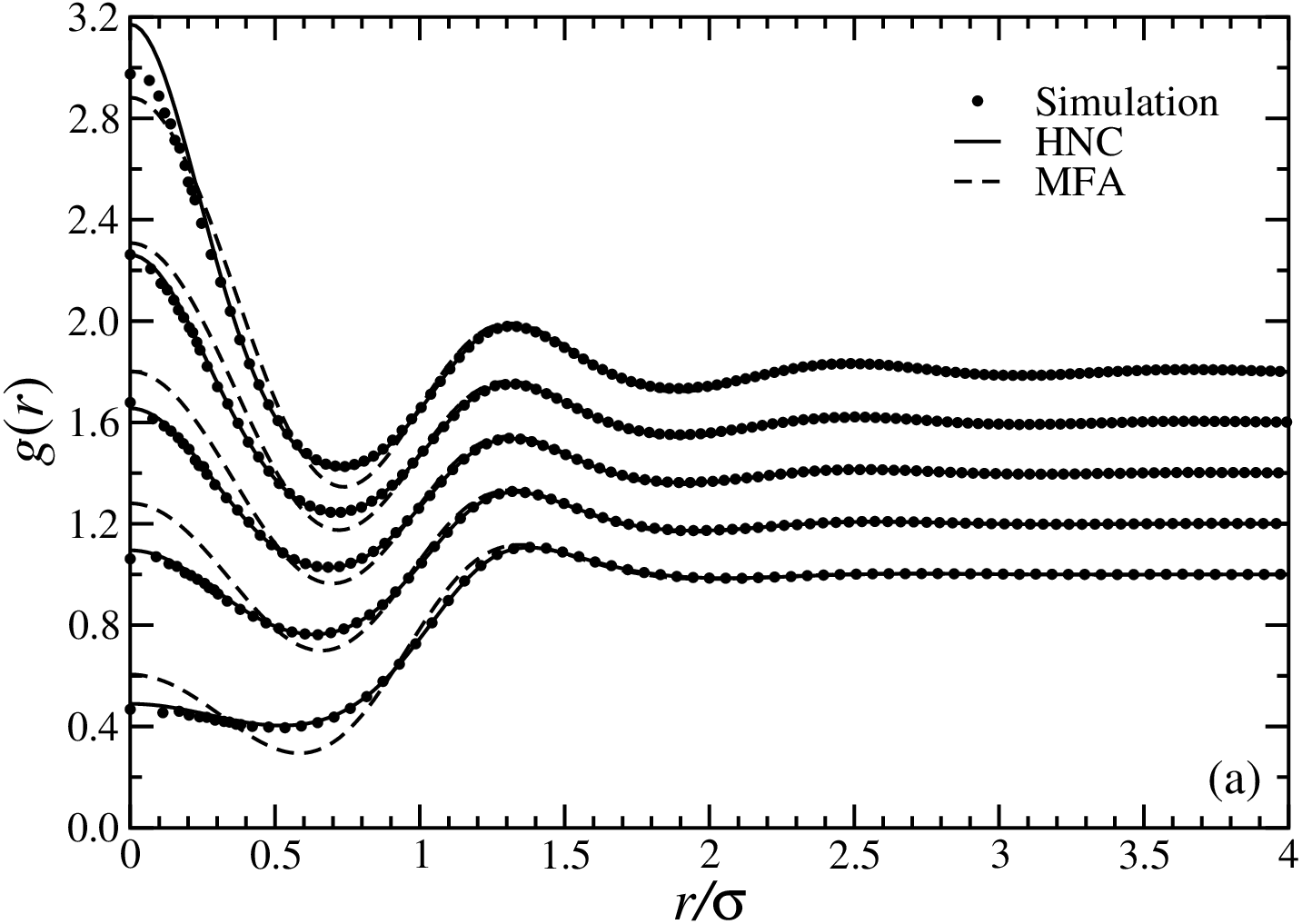}
      \end{minipage}
      \begin{minipage}[t]{8.3cm}
      \includegraphics[width=8.2cm, clip]
        {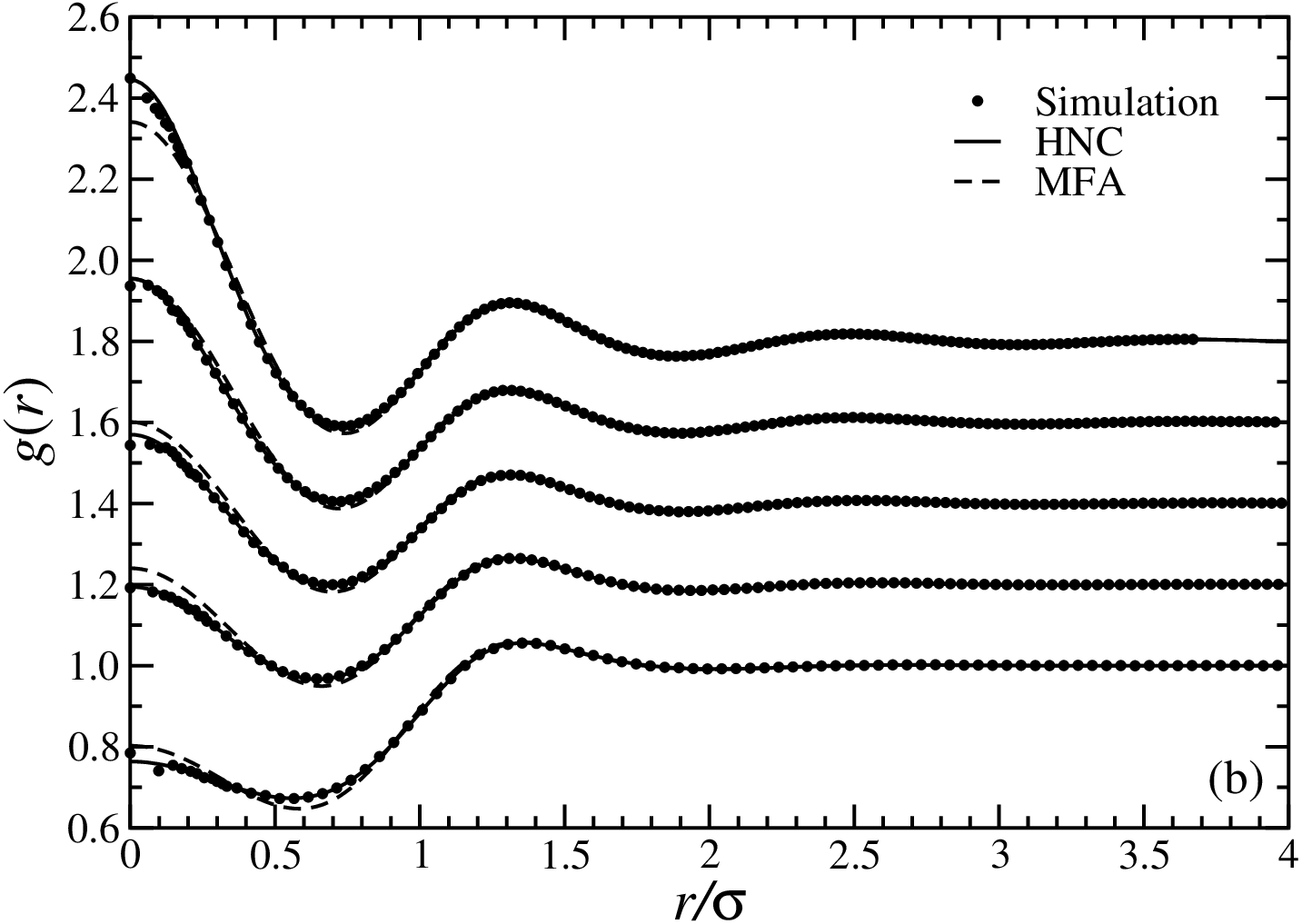}
      \end{minipage}
      \begin{minipage}[t]{8.3cm}
      \includegraphics[width=8.2cm, clip]
        {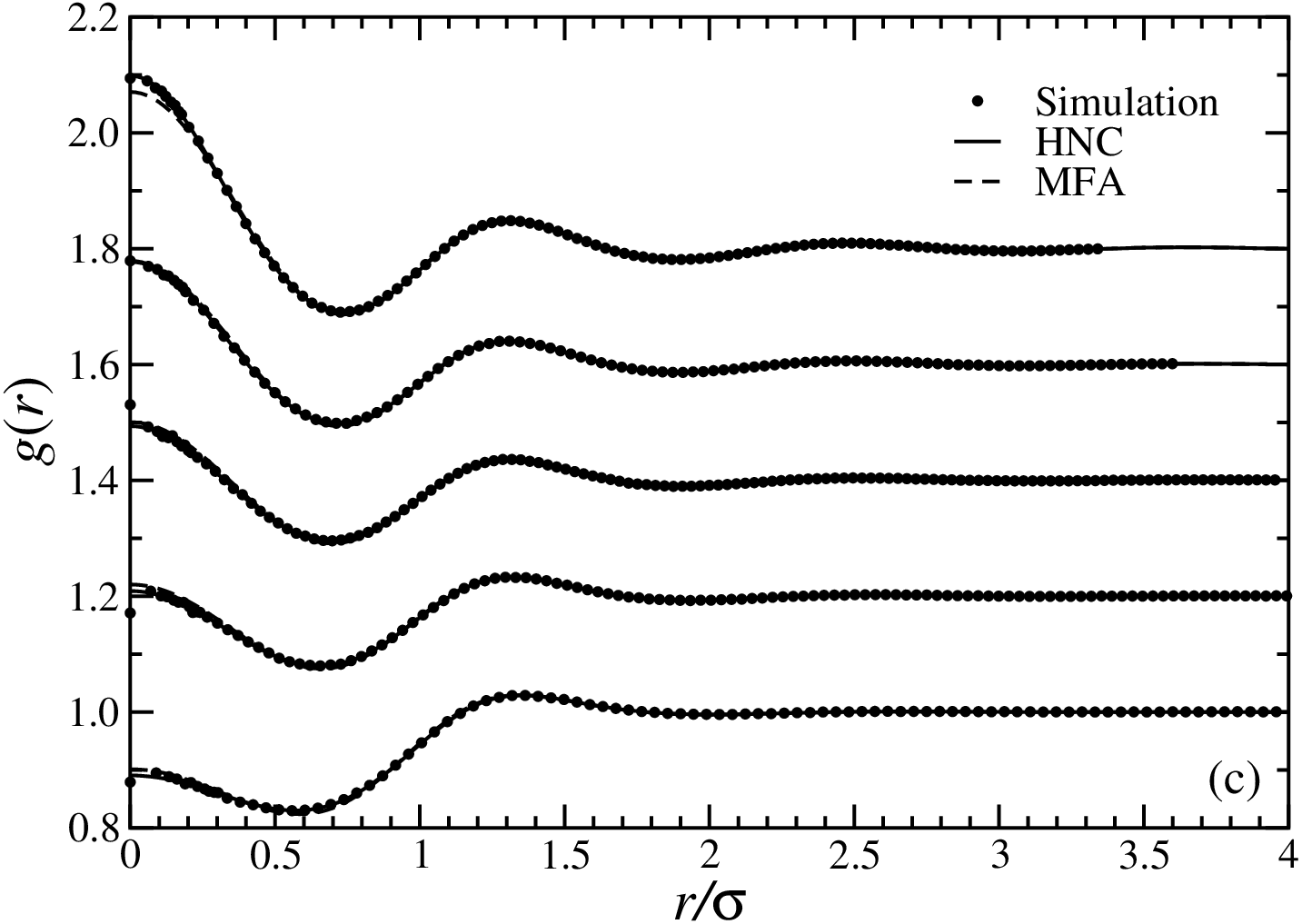}
      \end{minipage}
      \end{center}
\caption{Radial distribution functions $g(r)$ of the GEM-4 model
as obtained by Monte Carlo simulation (points), the HNC-closure
(solid lines) and the MFA (dashed lines), at various temperatures
and densities. For clarity, the curves on every panel have been
shifted upwards by certain amounts, which are given below in
square brackets, following
the value indicating the density $\rho^{*}$.
(a) $T^{*} = 0.5$ and densities, from bottom to top:
$\rho^{*} = 0.5$ [0]; $\rho^{*} = 1.0$ [0.2]; $\rho^{*} = 1.5$ [0.4];
$\rho^{*} = 2.0$ [0.6]; $\rho^{*} = 2.5$ [0.8].
(b) $T^{*} = 1.0$ and densities, from bottom to top:
$\rho^{*} = 1.0$ [0]; $\rho^{*} = 2.0$ [0.2]; $\rho^{*} = 3.0$ [0.4];
$\rho^{*} = 4.0$ [0.6]; $\rho^{*} = 5.0$ [0.8].
(c) $T^{*} = 2.0$ and densities, from bottom to top:
$\rho^{*} = 2.0$ [0]; $\rho^{*} = 4.0$ [0.2]; $\rho^{*} = 6.0$ [0.4];
$\rho^{*} = 8.0$ [0.6]; $\rho^{*} = 10.0$ [0.8].}
\label{gofr:fig}
\end{figure}

In Fig.\ \ref{gofr:fig} we show comparisons for the function 
$g(r)$ as obtained from the MC simulations and from the HNC closure
for a variety of temperatures and densities. It can be seen that
agreement between the two is obtained, to a degree of
quality that is excellent. Tiny deviations
between the HNC and MC results appear only at the highest density
and for a small region around $r = 0$ for 
low temperatures, $T^{*} = 0.5$. 
Otherwise, the system at hand
is described by the HNC with an extremely high accuracy  
and for a very broad range of temperatures and densities. 
We note that, although in Fig.\ \ref{gofr:fig} we restrict ourselves
to temperatures $T^{*} \leq 2.0$, the quality of the HNC remains
unaffected also at higher temperatures.\cite{criterion} 

In attempting to gain some insight into the remarkable ability of
the HNC to describe the fluid structure at such a high level of
accuracy,
it is useful to recast this
closure in density-functional language. Following the famous
Percus idea,\cite{hansen:book,percus:62,percus:64,denton:pra:91,yethiraj:jcp:01} 
the quantity $\rho g(r)$ can be identified
with the nonuniform density $\rho(r)$ of an inhomogeneous fluid that
results when a single particle is kept fixed at the origin, exerting
an `external' potential $V_{\rm ext}(r) = v(r)$ on the rest of the
system. Following standard procedures from density functional 
theory, we find that the sought-for density profile $\rho(r)$ 
is given by
\begin{equation}
\rho(r) = \Lambda^{-3}
\exp\left\{\beta\mu - \beta v(r) 
- \frac{\delta \beta F_{\rm ex}[\rho]}{\delta \rho(r)}\right\}, 
\label{profile_dft:eq}
\end{equation}
where $\Lambda$ is the thermal de Broglie wavelength and $\mu$
the chemical potential associated with average density $\rho$
and temperature $T$. Moreover, $F_{\rm ex}[\rho]$ is the
intrinsic excess free energy, a {\it unique} functional of the
density $\rho(r)$. As such, $F_{\rm ex}[\rho]$ can be expanded
in a functional Taylor series around its value for a uniform liquid
of some (arbitrary) reference density $\rho_0$. For the problem
at hand, $\rho_0 = \rho$ is a natural choice and we obtain\cite{singh}
\begin{eqnarray}
\nonumber
\beta F_{\rm ex}[\rho] = \beta F_{\rm ex}(\rho) 
& - & \sum_{n=1}^{\infty}\frac{1}{n!}\int\int\cdots\int
{\rm d}^3r_1{\rm d}^3r_2\ldots{\rm d}^3r_n
\\
\nonumber
&\times&
c_0^{(n)}({\bf r}_1,{\bf r}_2,\ldots,{\bf r}_n;\rho)
\\
&\times&
\Delta\rho({\bf r}_1)\Delta\rho({\bf r}_2)\ldots\Delta\rho({\bf r}_n), 
\label{taylor_expand:eq}
\end{eqnarray}
where $F_{\rm ex}(\rho)$ denotes the excess free energy of the 
{\it homogeneous} fluid, as opposed to that of the inhomogeneous
fluid, $F_{\rm ex}[\rho]$, and
\begin{equation}
\Delta\rho({\bf x}) \equiv \rho({\bf x}) - \rho.
\label{deltarho:eq}
\end{equation} 
The basis of the expansion given by Eq.\ (\ref{taylor_expand:eq}) above is
the fact that $F_{\rm ex}[\rho]$ is the generating functional for
the hierarchy of the direct correlation functions (dcf's) $c_0^{(n)}$.
In particular, $-\beta^{-1}c_0^{(n)}$ 
is the $n$-th functional derivative of the excess
free energy with respect to the density field, evaluated at the
uniform density $\rho$:\cite{singh,evans:78}
\begin{equation}
c_0^{(n)}({\bf r}_1,{\bf r}_2,\ldots,{\bf r}_n;\rho) =
-\frac{\delta^n \beta F_{\rm ex}[\rho]}
{\delta\rho({\bf r}_1)\delta\rho({\bf r}_2)\ldots
\delta\rho({\bf r}_n)}\Bigg|_{\rho(.) = \rho}.
\label{cofrn:eq}
\end{equation}
As the functional derivatives are evaluated for a uniform system,
translational and rotational invariance reduce the number of variables
on which the $n$-th order 
dcf's $c_0^{(n)}$ depend; in fact,
$c_0^{(1)}({\bf r}_1;\rho)$ is a position-independent constant and 
equals $-\beta\mu_{\rm ex}$, where $\mu_{\rm ex}$ is the excess 
chemical potential.\cite{evans:78} 
Similarly, $c_0^{(2)}({\bf r}_1,{\bf r}_2;\rho)$
is simply the 
Ornstein-Zernike direct correlation function $c(|{\bf r}_1-{\bf r}_2|)$ 
entering in Eqs.\ (\ref{oz:eq}) and (\ref{hnc:eq}) above. The HNC
closure
is equivalent to jointly solving Eqs.\ 
(\ref{oz:eq}) and (\ref{profile_dft:eq}) by employing
an approximate excess free energy functional $F_{\rm ex}[\rho]$,
arising by a truncation of the expansion of Eq.\ (\ref{taylor_expand:eq}) 
at $n = 2$, i.e., discarding all terms with $n \geq 3$; this
is the famous Ramakrishnan-Yussouff second-order 
approximation.\cite{ry:79,haymet:81}
Indeed,
the so-called bridge function $b(r)$ can be written as an
expansion over integrals involving as kernels all the
$c_0^{(n)}$ with $n \geq 3$ and the HNC amounts to setting
the bridge function equal to zero.\cite{hansen:book,denton:pra:91,denton:pra:90} 

Whereas in many cases, such as the
one-component plasma,\cite{ocp1,ocp2} 
and other systems with long-range interactions, 
the
HNC is simply an adequate or, at best, a very good approximation,
for the case at hand the degree of agreement between the HNC and
simulation is indeed extremely high.
What is particularly important is that the accuracy of the HNC persists
for a very wide range of densities, at all temperatures considered.
This fact has far-reaching consequences, because it means that the
corresponding profiles $\rho({\bf x})$ that enter the 
multiple integrals in Eq.\ (\ref{taylor_expand:eq}) vary enormously
depending on the uniform density considered. 
Thus, it is tempting to conjecture that for the systems under
consideration (soft, penetrable particles at $T^{*} \gtrsim 1$),
not simply the integrals with $n \geq 3$ vanish but rather the
kernels themselves. In other words,
\begin{equation}
c_0^{(n)}({\bf r}_1,{\bf r}_2,\ldots,{\bf r}_n;\rho) \cong 0,
\qquad (n \geq 3).
\label{approx:eq}
\end{equation} 

\begin{figure}
      \begin{center}
      \begin{minipage}[t]{8.3cm}
      \includegraphics[width=8.2cm, clip]
        {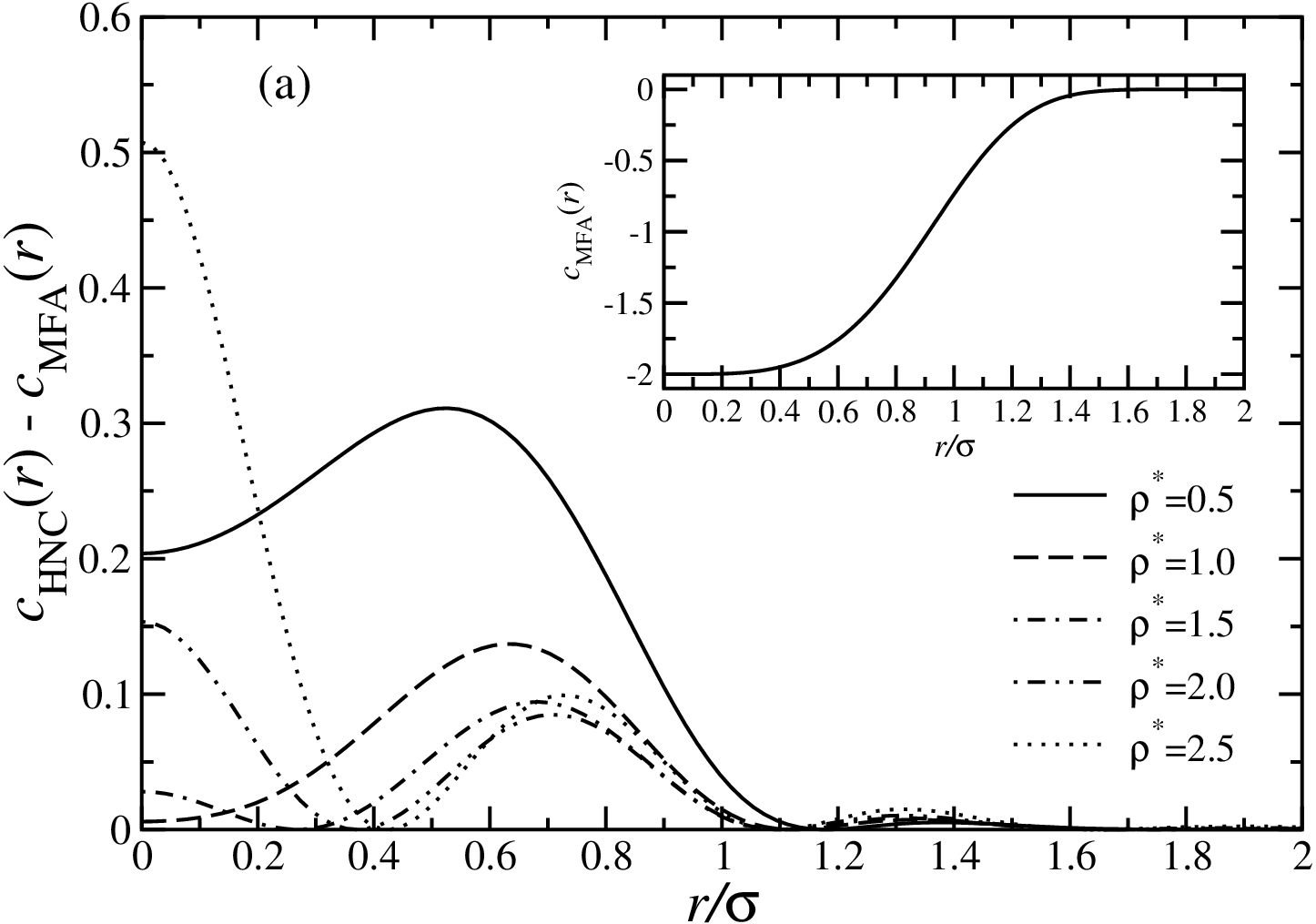}
      \end{minipage}
      \begin{minipage}[t]{8.3cm}
      \includegraphics[width=8.2cm, clip]
        {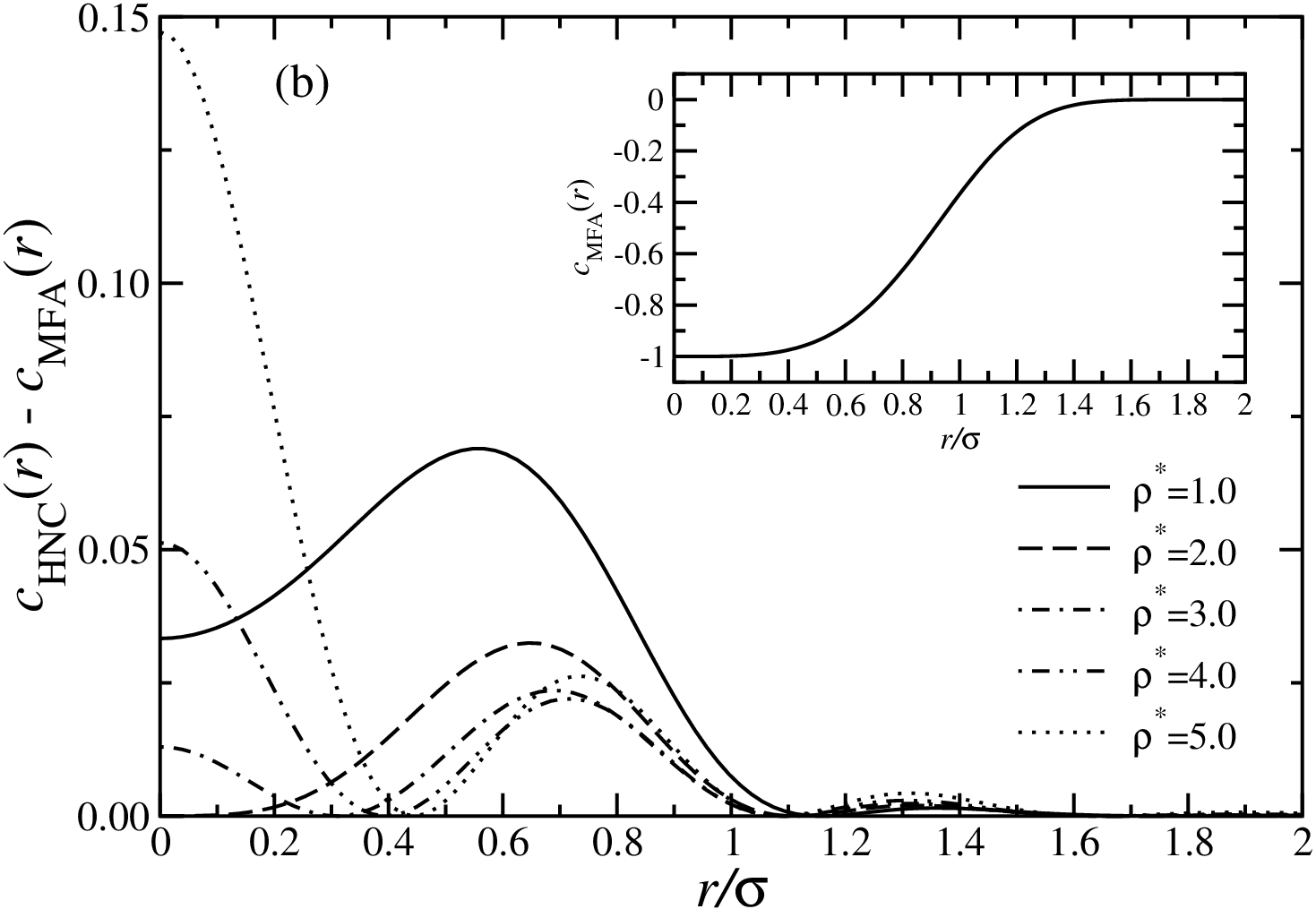}
      \end{minipage}
      \begin{minipage}[t]{8.3cm}
      \includegraphics[width=8.2cm, clip]
        {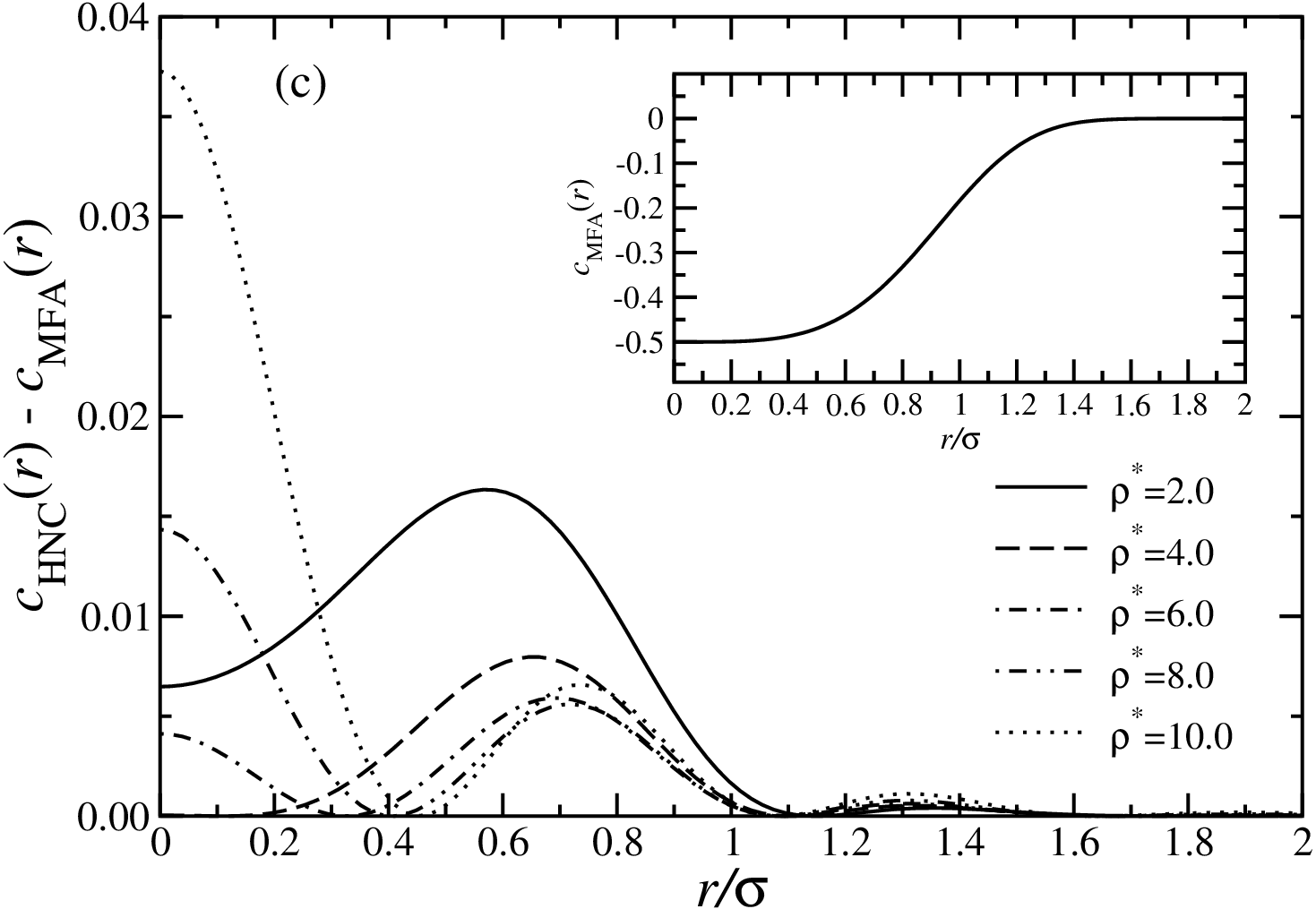}
      \end{minipage}
      \end{center}
\caption{The main plots show the difference between the 
direct correlation function $c(r)$ calculated in the HNC and its
MFA-approximation, $c(r) = -\beta v(r)$, for a GEM-4 model,
at various densities indicated in the legends.
The insets show the MFA approximation for the same quantity,
which is density-independent. Each panel 
corresponds to a different temperature: (a) $T^{*} = 0.5$;
(b) $T^{*} = 1.0$; (c) $T^{*} = 2.0$. These are exactly
the same parameter combinations as the ones for which
$g(r)$ is shown in Fig.\ \ref{gofr:fig}.}
\label{cofr:fig}
\end{figure}

The behavior of the higher-order dcf's
is related to the density-derivative of lower-order ones through
certain sum rules, to be discussed below. Hence, it is pertinent
to examine the density dependence of the dcf $c(r)$
of the HNC.
In Fig.\ \ref{cofr:fig} we show the difference between
the dcf $c_{\rm HNC}(r)$ and the 
mean-field approximation (MFA) to the same quantity:
\begin{equation}
c_{\rm MFA}(r) = -\beta v(r).
\label{cmfa:eq}
\end{equation} 
Eq.\ (\ref{cmfa:eq}) above is {\it meaningless} if the pair potential
$v(r)$ diverges as $r \to 0$, because $c(r)$ has to remain finite
at all $r$, as follows from exact diagrammatic expansions of the
same.\cite{hansen:book} In fact, the form $c(r) \sim -\beta v(r)$ denotes
the large-$r$ asymptotic behavior of $c(r)$. 
In our case, however, where
$v(r)$ lacks a hard core, the MFA-form for $c(r)$ cannot be 
a priori rejected on fundamental grounds; 
the quantity $-\beta v(r)$ remains bounded as $r \to0$. In fact, as can be seen
in Fig.\ \ref{cofr:fig}, the deviations between the MFA 
and the HNC-closure are very small for $T^* \gtrsim 1$.
In addition, the differences between $c_{\rm HNC}(r)$ and
$c_{\rm MFA}(r)$ drop,
both in absolute and in 
relative terms, as temperature grows, see the trend in 
Figs.\ \ref{cofr:fig}(a)-(c). At fixed temperature, the evolution
of the difference with density is nonmonotonic: it first drops
as density grows and then it starts growing again at the highest
densities shown in the three panels of Fig.\ \ref{cofr:fig}. 

Motivated by these findings, we
employ now a second closure, namely the above-mentioned
MFA, Eq.\ (\ref{cmfa:eq}). Introducing the latter into the
Ornstein-Zernike relation, Eq.\ (\ref{oz:eq}), we obtain the
MFA-results for the radial distribution function $g(r)$ that
are also shown in Fig.\ \ref{gofr:fig} with dashed lines. Before
proceeding to a critical comparison between the $g(r)$ obtained
from the two closures, it is useful to make a clear connection
between the MFA and the HNC.

As mentioned above,
the members of the sequence of the $n$-th order direct correlation functions 
are not independent from one another; rather, they are constrained 
to satisfy a corresponding hierarchy of sum rules, 
namely:\cite{denton:pra:91,barrat:prl:87,barrat:molphys:88,denton:pra:89}
\begin{eqnarray}
\nonumber
\int{\rm d}^3r_k c_0^{(n+1)}({\bf r}_1,\ldots,{\bf r}_{k-1},{\bf r}_{k},
{\bf r}_{k+1},\ldots,{\bf r}_{n+1};\rho) =
\\
\frac{\partial c_0^{(n)}({\bf r}_1,\ldots,{\bf r}_{k-1},
{\bf r}_{k+1},\ldots,{\bf r}_{n+1};\rho)}{\partial\rho}.
\label{inductive:eq}
\end{eqnarray}
In particular, for $n=2$, we have
\begin{eqnarray}
\nonumber
\int{\rm d}^3r' c_0^{(3)}(r,r',|{\bf r} - {\bf r'}|;\rho) & = &
\\
\nonumber
\int{\rm d}^3r' c_0^{(3)}(r',r,|{\bf r'} - {\bf r}|;\rho) & = &
\\
\frac{\partial c(r;\rho)}{\partial \rho},
\label{sumrule:eq}
\end{eqnarray}
where we have used the translational and
rotational invariance of the fluid phase to reduce the
number of arguments of the dcf's and we
show explicitly the generic dependence of $c(r)$ on $\rho$.
In the 
MFA, one assumes $c(r) = -\beta v(r)$, with the immediate consequence
\begin{equation}
\frac{\partial c(r)}{\partial \rho} = 0.
\label{deriv:eq}
\end{equation}
Eqs.\ (\ref{sumrule:eq}) and (\ref{deriv:eq}) imply that the integral
of $c_0^{(3)}$ with respect to any of its arguments must {\it vanish}
for {\it arbitrary} density. As $c_0^{(3)}$ has a complex dependence on
its arguments, this is a strong indication that $c_0^{(3)}$ itself
vanishes. In fact, both for the Barrat-Hansen-Pastore factorization
approximation for this quantity\cite{barrat:prl:87,barrat:molphys:88} 
and for the alternative,
Denton-Ashcroft $k$-space factorization 
of the same,\cite{denton:pra:89} the
vanishing of the right-hand side of Eq.\ (\ref{sumrule:eq}) implies that 
$c_0^{(3)} = 0$. Now, if $c_0^{(3)}$ vanishes, so does also its
density derivative and use of sum rule (\ref{inductive:eq}) for
$n=3$ implies $c_0^{(4)} = 0$. Successive use of the same for
higher $n$-values leads then to the conclusion that in the MFA:
\begin{equation}
c_0^{(n)}({\bf r}_1,{\bf r}_2,\ldots,{\bf r}_n;\rho) = 0,\qquad (n\geq 3).
\label{kill:eq}
\end{equation} 

We can now see that the accuracy of the HNC stems from the fact
that for these systems we can write
\begin{equation}
c(r) = -\beta v(r) + \varepsilon(r;\rho),
\label{epsilon:eq}
\end{equation}
where $\varepsilon(r;\rho)$ is a small function at all densities $\rho$,
offering concomitantly a very small, and {\it the only}, contribution
to the quantity $\partial c(r)/\partial\rho$. This implies that
$c_0^{(3)}$ itself is negligible by means of Eq.\ (\ref{sumrule:eq}).
Repeated use of Eq.\ (\ref{inductive:eq}) leads then to 
Eqs.\ (\ref{approx:eq}) and shows that the contributions from the
$n > 2$ terms, that are ignored in the HNC, are indeed negligible.
The HNC is, thus, very
accurate, due to the strong mean-field character
of the fluids at hand.\cite{likos:pusey} The deviations between the HNC and the MFA
come through the function $\varepsilon(r;\rho)$ above.

Let us now return to the discussion of the results for $g(r)$ and
$c(r)$ and the relative quality of the two closures at various
thermodynamic points. Referring first to Fig.\ \ref{gofr:fig}(c), we
see that at $T^{*} = 2.0$ both the HNC and the MFA perform 
equally well.
The agreement between the two (and between the MFA and MC)
worsens somewhat at $T^{*} = 1.0$ and even more at $T^{*} = 0.5$. The
MFA is, thus, an approximation valid for $T^{*} \gtrsim 1$, in agreement with
previous results.\cite{criterion} The reason lies in the
accuracy of the low-density limit of the MFA. In general, as
$\rho\to 0$, $c(r)$ tends to the Mayer
function $f(r) = \exp[-\beta v(r)] - 1$. If $T^{*} \gtrsim 1$, one
may expand the exponential to linear order and obtain
$f(r) \cong -\beta v(r)$, so that the MFA can be fulfilled.

In the HNC, it is implicitly assumed that all dcf's
with $n \geq 3$ vanish. In the MFA, this is also the case. The two
closures differ in one important point, though: in the HNC, the
second-order direct correlation function is {\it not} prescribed
but rather {\it determined}, so that both the `test-particle equation',
Eq.\ (\ref{profile_dft:eq}), and the Ornstein-Zernike relation, 
Eq.\ (\ref{oz:eq}),
are fulfilled. In the MFA, it is a priori assumed that 
$c(r) = -\beta v(r)$, which is introduced into the Ornstein-Zernike
relation and thus $h(r)$ is determined. This is one particular
way of obtaining $g(r)$ in the MFA, called the 
Ornstein-Zernike route. Alternatively, one could follow the
test-particle route in solving Eq.\ (\ref{profile_dft:eq}) in conjunction
with Eq.\ (\ref{taylor_expand:eq}) and the MFA-approximation,
Eq.\ (\ref{cmfa:eq}). In this case, the resulting expression for
the total correlation function $h(r)$ in the MFA 
reads as
\begin{equation}
h(r) = \exp[-\beta v(r) -\beta\rho(h*v)(r)] - 1,
\label{testpart:eq}
\end{equation}
with $*$ denoting the convolution. Previous studies
with ultrasoft systems have shown that the test-particle
$h(r)$ from the MFA is closer to the HNC-result than the 
MFA result obtained from the Ornstein-Zernike route.\cite{ingo:06} 
The discrepancy
between the two is a measure of the approximate character of the MFA;
were the theory to be exact, all routes would give the same result.
As a way to quantify the approximations involved in the MFA,
let us attempt to {\it impose} consistency between the test-particle
and Ornstein-Zernike routes. Since $c(r) = -\beta v(r)$ in this
closure, the exponent in Eq.\ (\ref{testpart:eq}) 
above is just the right-hand side
of the Ornstein-Zernike relation, Eq.\ (\ref{oz:eq}). Thus, if we
insist that the latter is fulfilled, we obtain the constraint
\begin{equation}
h(r) = \exp[h(r)] - 1,
\label{hofr:eq}
\end{equation}
which is strictly satisfied only for $h(r) = 0$. However, as long
as $|h(r)| < 1$, one can linearize the exponential and an identity
follows; the internal inconsistency of the MFA is of quadratic
order in $h(r)$ and it follows that 
the MFA provides an accurate closure for the
systems at hand, as long as $|h(r)|$ remains small. This explains
the deviations between MFA and MC seen at small $r$ for the
highest density at $T^{*} = 2.0$, Fig.\ \ref{gofr:fig}(c). The same 
effect 
can also be seen in Fig.\ \ref{cofr:fig}(c) as a growth of the
discrepancy between $c_{\rm HNC}(r)$ and $c_{\rm MFA}(r)$ at small
$r$-values for the highest density shown. In absolute terms,
however, this discrepancy remains very small. Note also that 
discrepancies at small $r$-values become strongly suppressed
upon taking a Fourier transform, due to the additional geometrical
$r^2$-factor involved in the three-dimensional integration.

It can therefore be seen that the MFA and the HNC are closely related
to one another: the HNC is so successful due to the strong mean-field
character of the systems under consideration. This fact has also
been established and extensively discussed for the case of the
Gaussian model,\cite{likos:gauss,louis:mfa} i.e., the $m=2$ member of the
GEM-$m$ class. Once more, the HNC and the MFA there are very accurate for
high densities and/or temperatures, where $h(r) \cong 0$ and the
system's behavior develops similarities with an `incompressible ideal gas',\cite{likos:gauss} 
in
full agreement with the remarks presented above. Subsequently,
the MFA and HNC closures have been also successfully applied to the
study of structure and thermodynamics of 
binary soft mixtures.\cite{louis:mfa,archer1,archer2,archer3,archer4,finken}

\begin{figure}
 \begin{center}
 \includegraphics[width=8.5cm, clip]
        {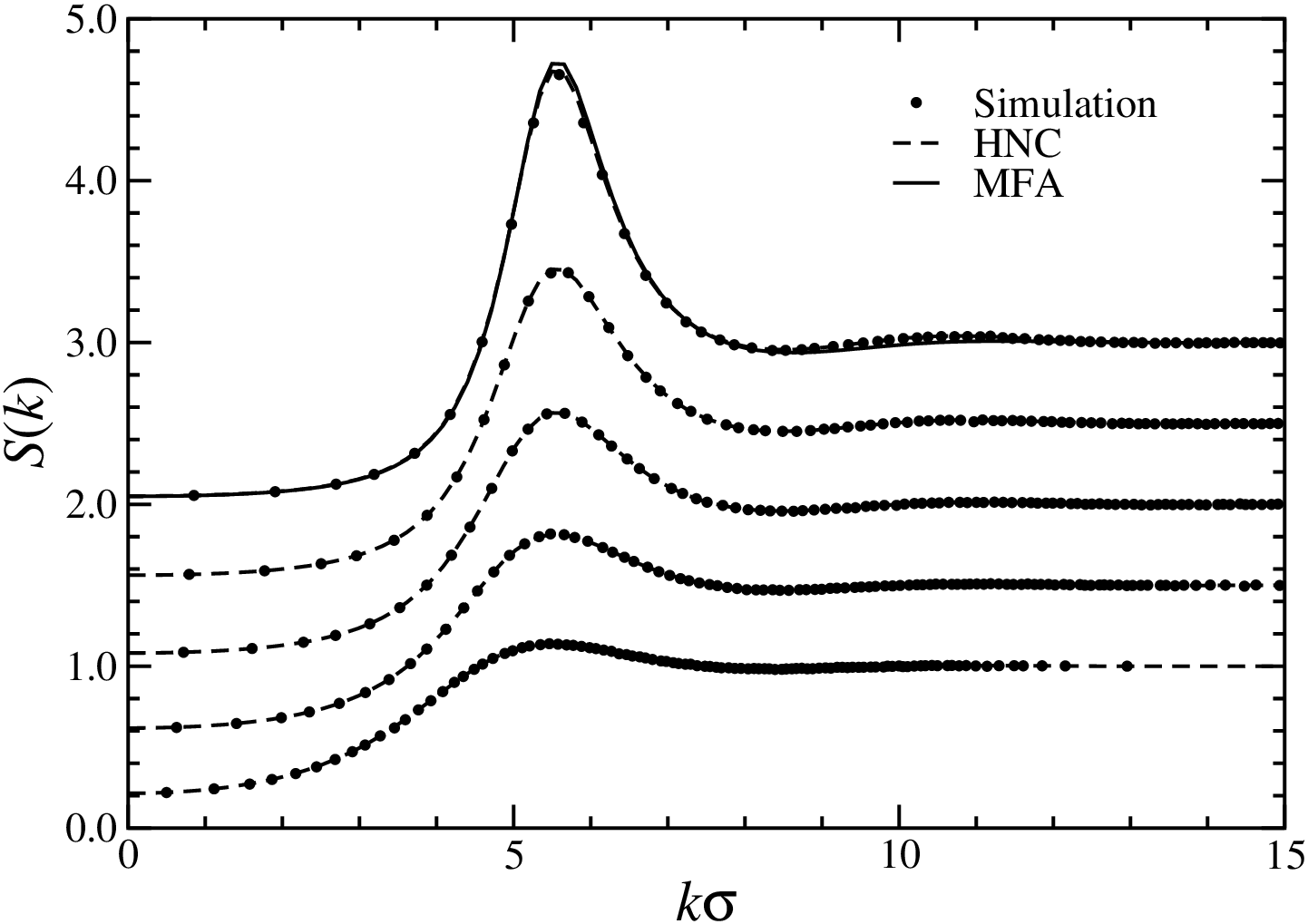}
 \end{center}
\caption{The structure factor $S(k)$ of the GEM-4 model at
temperature $T^{*} = 1.0$ and various densities, indicated
below. For clarity, the curves have been shifted vertically by
amounts shown in the square brackets, following the numbers
that indicate the values of the density $\rho^{*}$.
From bottom to top: $\rho^{*} = 1.0$ [0]; $\rho^{*} = 2.0$ [0.5];
$\rho^{*} = 3.0$ [1.0]; $\rho^{*} = 4.0$ [1.5]; $\rho^{*} = 5.0$ [2.0].
The points are results from Monte Carlo simulations and the
dashed lines from the HNC. As the HNC- and MFA-curves run very
close to each other, we show the MFA-result by the solid curve
only for the highest density, $\rho^{*} = 5.0$.}
\label{sofk:fig}
\end{figure}

A crucial
difference between the Gaussian model, which belongs to the
$Q^{+}$-class, and members of the $Q^{\pm}$-class, which are
the subject of the present work, lies in the consequences of the
MFA-closure on the structure factor $S(k)$ of the system. Since
$c(r) = -\beta v(r)$, the Ornstein-Zernike relation leads to
the expression
\begin{equation}
S(k) = \frac{1}{1+\beta\rho\tilde v(k)}.
\label{sofk:eq}
\end{equation}
Whereas for $Q^{+}$-potentials $S(k)$ is devoid of pronounced
peaks that exceed the asymptotic value $S(k\to\infty) = 1$, 
for $Q^{\pm}$-systems a local maximum of $S(k)$ appears at the value
$k_*$ for which $\tilde v(k)$ attains its negative minimum.
In Fig.\ \ref{sofk:fig} we show representative results for
the system at hand, where it can also be seen that the 
HNC and the MFA yield practically indistinguishable results.
In full agreement with the MC simulations, the location of the main peak of
$S(k)$ is {\it density-independent}, a feature unknown for
usual fluids, having its origin in the fact that $c(r)$ itself
is density independent.

Associated with this is the development of 
a $\lambda$-line,\cite{archer4,finken,stell,ciach}
also known as {\it Kirkwood instability},\cite{kirkwood:51}
on which the denominator 
in Eq.\ (\ref{sofk:eq}) vanishes at $k_*$ and 
thus $S(k_*) \to \infty$. The locus
of points $(\rho_{\lambda},T_{\lambda})$ on the 
density-temperature plane for
which this divergence takes place is, evidently, 
given by
\begin{equation}
\frac{T^*_{\lambda}}{\rho^*_{\lambda}} = -{\tilde \phi(k_*\sigma)}.  
\label{lambda:eq}
\end{equation}
In the region $\rho \geq \rho_{\lambda}$ 
(equivalently: $T \leq T_{\lambda}$) on the $(\rho,T)$-plane,
the MFA predicts that the fluid is absolutely unstable, since
the structure factor there has multiple divergences and also develops
negative parts. This holds {\it only} for $Q^{\pm}$-systems; for
$Q^{+}$-ones the very same line of argumentation leads to the
opposite conclusion, namely that the fluid is the phase of stability
at high densities and/or temperatures. The latter conclusion has already
been reached by Stillinger and
coworkers\cite{still1,still2,still3,still4,still5} 
in their pioneering work of the
Gaussian model in the mid-1970's, and explicitly
confirmed by extensive theory and computer simulations many
years later.\cite{stillinger:physica,likos:gauss,saija1, saija2, saija3, archer}
However, Stillinger's original argument was based on duality relations
that are strictly fulfilled only for the Gaussian model, whereas
the MFA-arguments are quite general.

\subsection{Nonuniform fluids}

Having established the validity of the MFA for vast domains in the
phase diagram of the systems under consideration as far as the
{\it uniform} fluids are concerned, we now turn our attention
to nonuniform ones. Apart from an obvious general interest in the
properties of nonuniform fluids, the necessity to consider 
deviations from homogeneity in the density for $Q^{\pm}$-models
is dictated by the $\lambda$-instability mentioned above: the
theory of the uniform fluid
contains its own breakdown, thus the system
has to undergo a phase transformation to a phase with a 
spontaneously broken translational symmetry. Whether this
transformation takes place {\it exactly} on the instability line
or already at densities $\rho < \rho_{\lambda}$ 
(or temperatures $T > T_{\lambda}$) and which is
the stable phase are some of the questions that have to be 
addressed. Density-functional theory of inhomogeneous systems
is the appropriate theoretical tool in this direction.

Let us consider a path $\rho_{\chi}({\bf r})$ in the space 
of density functions, which is characterized
by a single parameter $\chi$; this path starts at some reference
density $\rho_{\rm r}({\bf r})$ and terminates at another
density $\rho({\bf r})$. 
The uniqueness of the excess free energy functional and its
dependence on the inhomogeneous density field $\rho({\bf r})$
allow us to 
integrate 
$\partial F_{\rm ex}[\rho]/\partial\chi$
along this path,
obtaining $F_{\rm ex}[\rho]$, provided that
$F_{\rm ex}[\rho_{\rm r}]$ is known. A convenient
parametrization reads as
$\rho_{\chi}({\bf r}) = \rho_{\rm r}({\bf r})
+\chi[\rho({\bf r})-\rho_{\rm r}({\bf r})]$,
with $\chi = 0$ corresponding to
$\rho_{\rm r}({\bf r})$ and $\chi = 1$ to $\rho({\bf r})$.
The excess free energy of the final
state can be expressed as\cite{evans:78} 
\begin{eqnarray}
\nonumber
\beta F_{\rm ex}[\rho]  & = & 
\beta F_{\rm ex}[\rho_{\rm r}] 
\\
 & - &  \int_0^{1} {\rm d}\chi
\int{\rm d}^3r c^{(1)}({\bf r};[\rho_{\chi}])
\Delta\rho({\bf r}),
\label{chi_path:eq}
\end{eqnarray}
where $c^{(1)}({\bf r};[\rho_{\chi}])$ denotes the first 
functional derivative of the quantity $-\beta F_{\rm ex}[\rho]$
evaluated at the inhomogeneous density $\rho_{\chi}({\bf r})$,
and $\Delta\rho({\bf r}) = \rho({\bf r})-\rho_{\rm r}({\bf r})$.
Since $c^{(1)}({\bf r};[\rho_{\chi}])$
is in its own turn a unique functional of the density profile,
repeated use of the same argument leads to a functional Taylor
expansion of the excess free energy around that of a reference
system,
an expansion that extends to infinite order.
For the {\it particular} choice of a {\it uniform} reference system,
$\rho_{\rm r}({\bf r}) = \rho$,
we obtain then the Taylor series of Eq.\ (\ref{taylor_expand:eq}).
In general, however, the reference system does not have to have
the same average density as the final one, hence the uniform
density $\rho$ there must be replaced by a more general quantity
$\rho_0$. 

The usefulness of Eq.\ (\ref{taylor_expand:eq}) in calculating
the free energies of extremely nonuniform phases, such as crystals,
is limited both on principal and on practical grounds. Fundamentally,
there is {\it no small parameter} guiding such an expansion,
since the differences between the nonuniform density of a crystal
and that of a fluid are enormous; the former has extreme variations
between lattice- and interstitial regions. Hence, the very
convergence of the series is in doubt.\cite{singh} 
In practice, 
the direct correlation functions for $n = 3$ are very cumbersome to
calculate\cite{gerhard:00} and those for $n \geq 4$ are practically 
unknown.\cite{denton:pra:91} The solution is either to arbitrarily
terminate the series at second order\cite{ry:79} or to seek for
nonperturbative functionals.\cite{singh,likos:prl:92,likos:jcp:93,wda1,wda2,mwda,henderson}
In our case, however, things are
different because, for the systems we consider,
we have given evidence that the dcf's
of order $n > 2$ are extremely small and we take them
at this point as vanishing. Then,  
the functional Taylor
expansion of the free energy $F_{\rm ex}[\rho]$ terminates
(to the extent that the approximation holds) at second order.
The Taylor series becomes a finite sum and 
convergence is not an issue any more.

Let us, accordingly, expand $F_{\rm ex}[\rho]$
around an {\it arbitrary}, homogeneous reference fluid of density $\rho_0 = N_0/V$,
taking into account that the volume $V$ is fixed but the system
with density $\rho({\bf r})$ contains $N$ particles, whereas
the reference fluid contains $N_0$ particles and, in general,
$N \neq N_0$:
\begin{eqnarray}
\nonumber
\beta F_{\rm ex}[\rho]  & = &  \beta F_{\rm ex}(\rho_0) 
- c_0^{(1)}(\rho_0)\int{\rm d}^3r[\rho({\bf r}) - \rho_0] 
\\
\nonumber
& - &  
\frac{1}{2}\int\int{\rm d}^3r{\rm d}^3r' c_0^{(2)}(|{\bf r}-{\bf r'}|;\rho_0)
\\
& \times &
[\rho({\bf r}) - \rho_0][\rho({\bf r'}) - \rho_0].
\label{taylor:eq}
\end{eqnarray}

Using $c_0^{(2)}(r; \rho_0) =
  c_0^{(2)}(r) \equiv c(r) = -\beta v(r)$ and the
sum rule (\ref{inductive:eq}) for $n=1$ together with the vanishing
of the excess chemical potential at zero density, we readily obtain
\begin{equation}
c_0^{(1)}(\rho_0)  =  -\beta {\tilde v}(0) \rho_0.
\label{c1:eq}
\end{equation}
Formally substituting in Eq.\ (\ref{taylor:eq}),
$\rho_0 \to 0$ and $\rho({\bf r}) \to \rho_0$ and making use of
the fact that the excess free energy of a system vanishes with the density,
we also obtain the dependence
of the fluid excess free energy on the density:
\begin{equation}
\beta F_{\rm ex}(\rho_0)  =  \frac{N_0}{2}\beta{\tilde v}(0)\rho_0,
\label{frex_liq:eq}  
\end{equation}
with the particle number $N_0$ {\it of the reference fluid} and
the Fourier transform ${\tilde v}(k)$ of the interaction potential.
Introducing Eqs.\ (\ref{c1:eq}) and (\ref{frex_liq:eq})
into the Taylor expansion, Eq.\ (\ref{taylor:eq})
above, we obtain:
\begin{eqnarray}
\nonumber
\beta F_{\rm ex}[\rho] & = & \frac{N_0}{2}\beta{\tilde v}(0)\rho_0
\\
\nonumber 
& + & \beta{\tilde v}(0)\rho_0
(N - N_0) 
\\
\nonumber
& + & \frac{\beta}{2}\int\int{\rm d}^3r{\rm d}^3r'v(|{\bf r} - {\bf r'}|)
\rho({\bf r})\rho({\bf r'})
\\
\nonumber
& - & \beta\rho_0\int\int{\rm d}^3r{\rm d}^3r'v(|{\bf r} - {\bf r'}|)
\rho({\bf r})
\\
& + & \frac{N_0}{2}\beta{\tilde v}(0)\rho_0.
\label{expand:eq}
\end{eqnarray}
Introducing
${\bf x} \equiv {\bf r}-{\bf r'}$ the
fourth term above becomes:
\begin{equation}
- \beta\rho_0\int{\rm d}^3r\,\rho({\bf r})\int{\rm d}^3x\,v(|{\bf x}|)
 = -\beta N \rho_0 {\tilde v}(0).
\label{third:eq}
\end{equation}
Now the sum of the 1st, 2nd, 4th and 5th term in Eq.\ (\ref{expand:eq}) yields:
\begin{eqnarray}
\nonumber
\frac{N_0}{2}\beta{\tilde v}(0)\rho_0 + \beta{\tilde v}(0)\rho_0(N-N_0)
\\
\nonumber
-\beta N \rho_0 {\tilde v}(0) + \frac{N_0}{2}\beta{\tilde v}(0)\rho_0
\\
\nonumber
= \beta{\tilde v}(0)\left[\frac{N_0}{2}\rho_0 + \rho_0(N-N_0)
-N\rho_0 + \frac{N_0}{2}\rho_0\right]
\\
= \beta{\tilde v}(0)\left[N_0\rho_0 + \rho_0(N-N_0) - N\rho_0\right] = 0.
\end{eqnarray}
This is a remarkable cancelation because then only the 3rd term in
Eq. (\ref{expand:eq}) survives and we obtain:
\begin{equation}
F_{\rm ex}[\rho] = \frac{1}{2}\int\int{\rm d}^3r{\rm d}^3r'
v(|{\bf r} - {\bf r'}|)\rho({\bf r})\rho({\bf r'}),
\label{dftmfa:eq}
\end{equation}
which is our desired result.\cite{criterion,likos:gauss,louis:mfa} 

The derivation above 
demonstrates that the excess free energy
of {\it any} inhomogeneous phase 
for our ultrasoft fluids is given by Eq.\ (\ref{dftmfa:eq}),
irrespectively of the density of the reference fluid $\rho_0$.
This is particularly important because, usually, functional Taylor
expansions are carried out around a reference fluid whose density lies
close to the average density of the inhomogeneous system (crystal,
in our case). However, in our systems this is impossible. The
crystals occur predominantly in domains of the phase diagram in which the reference
fluid is meaningless, because they are on the high density 
side of the
$\lambda$-line. It is therefore important to be able to
justify the use of the functional and to avoid the inherent 
contradiction of expanding around an unstable fluid. 
In practice, of course, the higher-order dcf's do not 
exactly vanish,
hence deviations from result (\ref{dftmfa:eq}) are expected to occur,
in particular at low temperatures and densities. Nevertheless,
the comparisons with simulations, e.g., in  
Refs.\ [\onlinecite{criterion}], [\onlinecite{bianca:prl:06}] and [\onlinecite{archer4}] fully
justify our approximation.

A mathematical proof of the mean-field character for fluids
with infinitely long-range and infinitesimally strong
repulsions
has existed since the late 1970's, see
Refs.\ [\onlinecite{grewe1}] and [\onlinecite{grewe2}]. However, even far away
from fulfillment of this limit, and for conditions that are
quite realistic for soft matter systems, the mean-field 
behavior continues to be valid.\cite{criterion,likos:gauss,louis:mfa}
The mean-field result of Eq.\ (\ref{dftmfa:eq}) has been put
forward for the Gaussian model at high densities,\cite{likos:gauss}
on the basis of physical argumentation: in the absence of 
diverging excluded-volume interactions, at sufficiently high
densities any given particle sees an ocean of others -- the
classical mean-field picture. The mean-field character of the
Gaussian model for moderate to high temperatures was  
demonstrated independently in Ref.\ [\onlinecite{louis:mfa}].
Here, we have provided a more
rigorous justification of its validity, based on the vanishing
of high-order direct correlation functions in the fluid. 
It must also be noted that the mean-field approximation has
recently been applied to a system with a broad shoulder and
a much shorter hard-core interaction, providing good agreement
with simulation results\cite{primoz:07} and allowing for the
formulation of a generalized clustering criterion for 
the inhomogeneous phases. 

An astonishing similarity 
exists between the mean-field functional of Eq.\ (\ref{dftmfa:eq})
and an exact result derived for 
infinite-dimensional hard spheres. Indeed, for
this case Frisch 
{\it et al.}\cite{frisch:prl:85,frisch:jcp:86,frisch:pra:87,frisch:pre:99} as
well as Bagchi and Rice\cite{bagchi:jcp:88} have shown that
\begin{equation}
\beta F_{\rm ex}[\rho] = -\frac{1}{2}\int\int{\rm d}^D r{\rm d}^D r'
f(|{\bf r} - {\bf r'}|)\rho({\bf r})\rho({\bf r'}),
\label{mayer:eq}
\end{equation}
where $D \to \infty$ and $f(|{\bf r} - {\bf r'}|)$ is the Mayer function
of the infinite dimensional hard spheres. Again, one has a bilinear
excess functional whose integration kernel does not depend on the
density; in this case, this is minus the (bounded) Mayer function whereas for
mean-field fluids, it is the interaction potential itself,
divided by the thermal energy $k_{\rm B}T$. In fact, the Mayer function
and the direct correlation function coincide for infinite-dimensional
systems and higher-order contributions vanish there as 
well,\cite{bagchi:jcp:88} making the analogy with our three-dimensional,
ultrasoft systems complete. Accordingly, infinite dimensional
hard spheres have an instability at some finite $k$
at the density
$\rho_{\star}$, given by $\rho_{\star}{\tilde f}(k) = 1$. This so-called
{\it Kirkwood instability}\cite{kirkwood:51} 
is of the same nature as our $\lambda$-line
but hard hyperspheres are athermal, so it occurs at a single
point on the density axis and not a line on the density-temperature
plane. 
Following Kirkwood's work,\cite{kirkwood:51} it was
therefore argued\cite{frisch:pra:87,bagchi:jcp:88} that hard hyperspheres
might have a second-order freezing transition at the
density $\rho_{\star}$ expressed as
\begin{equation}
\rho_{\star} = 0.239(e/8)^{D/2}\exp[z_0(D/2)^{1/3}]D^{1/6},
\label{rhostar:eq}
\end{equation}
where the limit $D \to \infty$ must be taken and $z_0 = 1.8558$ is the
value of the minimum of the Bessel function $J_{D/2}(z)$ as $D \to \infty$.
Note that $\rho_{\star} \to 0$ as $D \to \infty$.
Later on, Frisch and Percus argued that
most likely the Kirkwood instability is never encountered
because it is preempted by a first-order freezing
transition.\cite{frisch:pre:99} 
In what follows, we will show analytically that this 
is also the case for our systems, which might provide a finite-dimensional
realization of the above-mentioned mathematical limit. 

\section{Analytical calculation of the freezing properties}
\label{analytical:sec}

As mentioned above, an obvious candidate for a spatially modulated 
phase is a periodic crystal. The purpose of this section is
to employ density functional theory in order to calculate the
freezing properties. Under some weak, simplifying assumptions,
the problem can be solved analytically. 

Adding the ideal contribution to the
excess functional of Eq.\ (\ref{dftmfa:eq}),
the free energy of any spatially modulated phase is obtained as
\begin{eqnarray}
\nonumber
F[\rho] & = & F_{\rm id}[\rho] + F_{\rm ex}[\rho]
\\
\nonumber
& = &  k_{\rm B}T\int{\rm d}^3r\left[\ln\left[\rho({\bf r})\Lambda^3\right]
-1\right] 
\\
& + & \frac{1}{2}\int\int{\rm d}^3r{\rm d}^3r'
v(|{\bf r}-{\bf r'}|)\rho({\bf r})\rho({\bf r'}).
\label{totalfren:eq}
\end{eqnarray} 
As we are interested in crystalline phases, we parametrize the
density profile as a sum of Gaussians centered around the lattice
sites $\{{\bf R}\}$, forming a Bravais lattice. In sharp contrast
with systems interacting by hard, diverging potentials, however,
the assumption of one particle per lattice site has to be dropped.
Indeed, it will be seen that $Q^{\pm}$ systems employ the strategy
of optimizing their lattice constant by adjusting the number
of particles per lattice site, $n_c$, at any given density 
$\rho$ and temperature $T$. Accordingly, we normalize the
profiles to $n_c$ and write
\begin{eqnarray}
\nonumber
\rho({\bf r}) & = &  n_c\left(\frac{\alpha}{\pi}\right)^{3/2}
\sum_{{\bf R}}e^{-\alpha\left({\bf r}-{\bf R}\right)^2}
\\
& = & \sum_{{\bf R}}\rho_l({\bf r}-{\bf R}),
\label{gaussian_real:eq}
\end{eqnarray}
where the occupation variable $n_c$ and the 
localization parameter $\alpha$ have to be determined variationally,
and the lattice site density $\rho_l({\bf r})$ is expressed as
\begin{equation}
\rho_l({\bf r}) = n_c\left(\frac{\alpha}{\pi}\right)^{3/2}
e^{-\alpha r^2}.
\label{orbital:eq}
\end{equation}
Contrary to crystals of single occupancy, thus, the number of
particles $N$ and the number of sites $N_s$ of the Bravais lattice
do not coincide. In particular, it holds
\begin{equation}
\frac{N}{N_s} = n_c,
\label{nc:eq}
\end{equation}
and we are interested in multiple site occupancies, i.e., $n_c > 1$ or
even $n_c \gg 1$; it will be shown that this clustering scenario indeed
minimizes the crystal's free energy.

It is advantageous, at this point, to express the periodic
density profile of Eq.\ (\ref{gaussian_real:eq}) as a Fourier series,
introducing the Fourier components $\rho_{\bf K}$ of the same:
\begin{equation}
\rho({\bf r}) = \sum_{{\bf K}}
e^{{\rm i}{\bf K}\cdot{\bf r}}\rho_{\bf K},
\label{denfour:eq}
\end{equation}
where the set $\{{\bf K}\}$ contains all reciprocal lattice vectors 
(RLVs) of the Bravais lattice formed by the set $\{{\bf R}\}$.
Accordingly, the inverse of (\ref{denfour:eq}) reads as:\cite{ashcroft}
\begin{eqnarray}
\nonumber
\rho_{\bf K} & = &  \frac{1}{v_c}\int_{{\mathcal C}}
{\rm d}^3r e^{{\rm i}{\bf K}\cdot{\bf r}}\rho({\bf r})
\\
& = & \frac{1}{v_c}
\int {\rm d}^3r e^{{\rm i}{\bf K}\cdot{\bf r}}\rho_l({\bf r}),
\label{deninfour:eq}
\end{eqnarray}
where the first integral extends over the elementary
unit cell ${\mathcal C}$ of the crystal and $v_c = V/N_s$ is the 
volume of ${\mathcal C}$, containing a single lattice site.
The second integral extends over all space, where 
use of the periodicity of $\rho({\bf r})$ and
its expression as a sum over lattice site densities, 
Eq.\ (\ref{gaussian_real:eq}), has been made. Using 
Eq.\ (\ref{orbital:eq}) we obtain
\begin{eqnarray}
\nonumber
\rho_{\bf K} & = & \frac{n_c}{v_c}e^{-K^2/(4\alpha)}
\\
& = & \rho e^{-K^2/(4\alpha)}.
\label{rhok:eq}
\end{eqnarray}
Note that the site occupancy $n_c$ does not appear explicitly
in the functional form of the Fourier components of $\rho({\bf r})$,
a feature that may seem paradoxical at first sight. However, for
fixed density $\rho$ and any crystal type, the lattice constant
and thus also the reciprocal lattice vectors ${\bf K}$ are affected
by the possibility of clustering, thus the dependence on $n_c$ 
remains, albeit in an implicit fashion. With the density being
expressed in reciprocal space, the excess free energy takes a
simple form that reads as
\begin{equation}
\frac{F_{\rm ex}}{N} = \frac{\rho}{2}\sum_{{\bf K}}\tilde v(K) 
e^{-K^2/(2\alpha)}.
\label{fexk:eq}
\end{equation}

The ideal term, $F_{\rm id}[\rho]$, can also be approximated
analytically, provided that the Gaussians centered
at different lattice sites do not overlap. Let $a$ denote the
lattice constant of any particular crystal. Then, for 
$\alpha a^2 \gg 1$, the ideal free energy of the crystal takes
the form
\begin{equation}
\frac{\beta F_{\rm id}}{N} = 
\ln n_c + \frac{3}{2}\ln\left(\frac{\alpha\sigma^2}{\pi}\right)
-\frac{5}{2} + 3\ln\left(\frac{\Lambda}{\sigma}\right),
\label{fida:eq}
\end{equation}
where the trivial last term will be dropped in what follows, since
is also appears in the expression of the free energy of the fluid
and does not affect any phase boundaries. Putting together 
Eqs.\ (\ref{fexk:eq}) and (\ref{fida:eq}), we obtain a variational
free energy per particle, $\tilde f$, for the crystal, that reads as
\begin{eqnarray}
\nonumber
\frac{F_{\rm id}+F_{\rm ex}}{N\epsilon} & \equiv &
\tilde f(n_c,\alpha^*;T^*,\rho^*) 
\\
\nonumber
& = &  
T^*\left[\ln n_c + \frac{3}{2}\ln\left(\frac{\alpha^*}{\pi}\right)
-\frac{5}{2}\right] 
\\
& + & \frac{\rho^*}{2}\sum_{\bf Y}\tilde\phi(Y)
 e^{-Y^2/(2\alpha^*)},
\label{fren_var:eq}
\end{eqnarray}
where $\alpha^* \equiv \alpha\sigma^2$ and ${\bf Y} \equiv {\bf K}\sigma$.
In the list of arguments of $\tilde f$ the first two are
variational parameters whereas the last two denote simply its
dependence on temperature and density. The free energy per particle,
$f_{\rm sol}(T^*,\rho^*) \equiv F/(N\epsilon)$ of the crystal is obtained
by minimization of $\tilde f$, i.e.,
\begin{equation}
f_{\rm sol}(T^*,\rho^*) = \min_{\{n_c,\alpha^*\}}
\tilde f(n_c,\alpha^*;T^*,\rho^*).
\label{minimize:eq}
\end{equation}

In carrying out the minimization, it proves
useful to measure the localization length of the Gaussian
profile, $\ell \equiv 1/\sqrt{\alpha}$, in units of the
lattice constant $a$ instead of units of $\sigma$.
To perform this change, we first express the average 
density $\rho$ of the crystal in terms of $n_c$ and $a$ as
\begin{equation}
\rho = \frac{z n_c}{a^3},
\label{den_a:eq}
\end{equation}
where $z$ is a lattice-dependent numerical coefficient of order
unity. Introduce now the quantity 
$\alpha a^2 = \gamma^{-1}$. Using Eq.\ (\ref{den_a:eq}) above, 
we obtain
\begin{equation}
\alpha^* 
= \gamma^{-1}\left(\frac{\rho^*}{zn_c}\right)^{2/3}.
\label{amin_nc:eq}
\end{equation}
This change of variables is just a mathematical transformation that
simplifies the mathematics to follow; all results to be derived
maintain their validity also in the original representation.
For a further discussion of this point, see also Appendix B.

Next we make the simplifying approximation to
ignore in the sum over reciprocal lattice vectors on the right-hand-side
of Eq.\ (\ref{fren_var:eq}) above all the RLVs beyond the first
shell, whose length is $Y_1 = K_1\sigma$. This is justified
already because of the
exponentially damping factors
$\exp[-Y^2/(2\alpha^*)]$ in the sum. In addition, the 
coefficients
$|\tilde\phi(Y)|$ themselves decay to
zero as $Y \to \infty$, with an asymptotic behavior that
depends on the form of $\phi(x)$ in real space. The 
length of the first
shell of RLVs of any Bravais lattice of lattice constant $a$
scales as $K_1 = \zeta/a$, with some positive, lattice-dependent
numerical constant $\zeta$ of order unity. Together with 
Eqs.\ (\ref{den_a:eq}) this implies that the length of the 
first RLV depends on the aggregation number $n_c$ as
\begin{equation}
Y_1(n_c) = \zeta\left(\frac{\rho^*}{z n_c}\right)^{1/3},
\label{y1nc:eq}
\end{equation}
and using Eq.\ (\ref{amin_nc:eq}), we see that the ratio $Y_1^2/(2\alpha^*)$
takes a form that depends {\it solely} on the parameter $\gamma$,
namely
\begin{equation}
\frac{Y_1^2(n_c)}{2\alpha^*} = \frac{\gamma\zeta^2}{2}.
\label{d:eq}
\end{equation}

Introducing Eqs.\ (\ref{amin_nc:eq}) and ({\ref{d:eq}}) into
(\ref{fren_var:eq}), we obtain another
functional form for the variational free energy,
$\bar f(n_c,\gamma;T^*,\rho^*)$, expressed
in the new variables. It can be seen that 
upon making the transformation
(\ref{amin_nc:eq}), 
the term 
$3/2\ln[(\alpha^*/\pi)]$ 
delivers a contribution {\it minus} $\ln n_c$
that exactly cancels the same term with a positive sign on
the right-hand side of Eq.\ (\ref{fren_var:eq}). 
Accordingly, the
{\it only} remaining quantity of the variational free energy
that still depends on $n_c$ is the length of the first 
nonvanishing RLV, $Y_1$, whose $n_c$ dependence is expressed
by Eq.\ (\ref{y1nc:eq}) above. 
Putting everything together,
we obtain  
\begin{eqnarray}
\nonumber
\bar f(n_c,\gamma;T^*,\rho^*) & = & T^*
\left[\ln \rho^* - 1 - \frac{3}{2}\left[\ln(\gamma\pi) - 1\right] - \ln z
\right] 
\\
\nonumber
& + &  \frac{1}{2}\rho^*\tilde\phi(0) 
\\
& + & \frac{\xi_1\rho^*}{2}\tilde\phi\left(Y_1(n_c)\right)e^{-\gamma\zeta^2/2},
\label{fbar:eq}
\end{eqnarray}
where $\xi_1$ is the coordination number of the {\it reciprocal} lattice.
Minimizing $\bar f$ with respect to $n_c$ is trivial and using
Eq.\ (\ref{d:eq}) we obtain
\begin{equation}
\frac{\partial {\bar f}}{\partial n_c} = 0
\Rightarrow \tilde \phi'(Y_1)Y_1^4 = 0,
\label{partialnc:eq}
\end{equation}
where the prime denotes the derivative with respect to the argument.
Evidently, $Y_1$ coincides with $y_*\equiv k_*\sigma$, the 
dimensionless wavenumber for which
the dimensionless Fourier transform of the interaction potential
attains its negative minimum. The other mathematical solution
of (\ref{partialnc:eq}), $Y_1 = 0$, can 
be rejected because it yields nonpositive second derivatives or,
on physical grounds, because it corresponds to a crystal with
$n_c \to \infty$,
whose occurrence would violate the thermodynamic stability
of the system.    
Regarding second derivatives,  
it can be easily shown that
\begin{equation}
\frac{\partial^2 \bar f}{\partial n_c^2}\Bigg|_{Y_1 = y_*} > 0, 
\label{second_nc:eq}
\end{equation}
and
\begin{equation}
\frac{\partial^2 \bar f}{\partial\gamma\partial n_c}\Bigg|_{Y_1 = y_*} = 0,
\label{second_ncgamma:eq}
\end{equation}
irrespective of $\gamma$. 

Having shown the coincidence of $Y_1$ with $y_*$, we set
$\tilde\phi(Y_1) = \tilde\phi(y_*) < 0$ in Eq.\ (\ref{fbar:eq})
above. Further, we notice that the term $T^*[\ln\rho^*-1]$ on the
right-hand side of Eq.\ (\ref{fbar:eq}) gives the
ideal free energy of a uniform fluid of density
$\rho^*$ and the term $\rho^*\tilde\phi(0)/2$ the excess part
of the same, see
Eq.\ (\ref{frex_liq:eq}). 
Subtracting, thus, the total fluid free
energy per particle, $f_{\rm fl}(T^*,\rho^*)$, we introduce
the difference $\Delta {\bar f} \equiv \bar f - f_{\rm fl}$,
which reads as
\begin{eqnarray}
\nonumber
\Delta \bar f(n_c(y_*),\gamma;T^*,\rho^*) = 
& - & \frac{3T^*}{2}\left[\ln(\gamma\pi) + 1 + \frac{2\ln z}{3}\right]
\\
& + & \frac{\xi_1\rho^*}{2}\tilde\phi(y_*)e^{-\gamma\zeta^2/2}. 
\label{deltaf:eq}
\end{eqnarray}

The requirement of no overlap between Gaussians
centered on different lattice sites restricts $\gamma$ to be small;
a very generous upper limit is $\gamma \leq 0.05$. For such
small values of $\gamma$, 
the first term on the right-hand side of Eq.\ (\ref{deltaf:eq}) above
is positive. 
This positivity
expresses the entropic cost
of localization that a crystal pays, compared to the fluid in which
the delocalized
particles possess translational entropy. This cost must be compensated
by a gain in the excess term, which is only possible if $\tilde\phi(y_*) < 0$.
An additional degree of freedom is offered by the candidate crystal
structures. 
The excess free energy is minimized by the
direct Bravais lattice whose reciprocal lattice has 
the maximum possible coordination number $\xi_1$.
The most highly coordinated periodic arrangement 
of sites is fcc, for which $\xi_1 = 12$. Therefore, 
in the framework of this approximation, the stable lattice
is bcc. It must be emphasized, though, that these results hold
as long as only the first shell of RLVs is kept in the excess
free energy. Inclusion of higher-order shells can, under
suitable thermodynamic conditions, stabilize fcc in favor of bcc.
We will return to this point later.

Choosing now $a$ as the edge-length of the {\it conventional}
lattice cell of the bcc-lattice, we have $z=2$ and
$\zeta = 2\sqrt{2}\pi$. Evidently,  
the lattice constant of the crystal is
density-independent, 
$a/\sigma = (2\sqrt{2}\pi)/y_*$, contrary to the case
of usual crystals, for which $a \propto \rho^{-1/3}$. 
The density-independence of $a$ is achieved
by the creation of clusters that consist of $n_c$ particles,
each of them occupying a  
lattice site. The proportionality
relation connecting $n_c$ and $\rho^*$ follows from
Eq.\ (\ref{y1nc:eq}) and reads as
\begin{equation}
n_c = \frac{8\sqrt{2}\pi^3}{y_*^3}\rho^*.
\label{proport:eq}
\end{equation}

It remains to minimize $\bar f$ (equivalently, $\Delta\bar f$)
with respect to $\gamma$ to determine the free energy of the crystal.
We are interested, in particular, in estimating the `freezing
line', determined by the equality of free energies of the fluid
and the solid.\cite{foot1}
Accordingly,
we search for the simultaneous solution of the equations
\begin{eqnarray}
\frac{\partial {\bar f}}{\partial \gamma} & = & 0,
\\
\Delta {\bar f} & = & 0,
\end{eqnarray}
resulting into
\begin{equation}
\frac{3T^*}{2\gamma} + 
\frac{\xi_1\zeta^2\rho^*}{4}\tilde\phi(y_*)e^{-\gamma\zeta^2/2}=0,
\label{dfdg:eq}
\end{equation}
and
\begin{equation}
\frac{\xi_1\rho^*}{2}\tilde\phi(y_*)e^{-\gamma\zeta^2/2}=
\frac{3T^*}{2}\left[\ln(\gamma\pi)+1+\frac{2\ln z}{3}\right].
\label{dfzero:eq}
\end{equation}
Substituting (\ref{dfzero:eq}) into (\ref{dfdg:eq}) and using $z = 2$ and
$\zeta = 2\sqrt{2}\pi$, we obtain an implicit equation for $\gamma$
that reads as
\begin{equation}
\gamma^{-1} = -4\pi^2
\left[\ln\left(\gamma\pi\right)+1+\frac{2\ln 2}{3}\right],
\label{gamma:eq}
\end{equation}
and has two solutions, $\gamma_1 \cong 0.018$ and $\gamma_2 \cong 0.038$.
Due to (\ref{second_nc:eq}) and (\ref{second_ncgamma:eq}),
the sign of the determinant of the Hessian matrix at the extremum
is set by the sign of $\partial^2{\bar f}/\partial \gamma^2$;
Using Eqs.\ (\ref{dfdg:eq}), (\ref{dfzero:eq}),
and (\ref{gamma:eq}) we obtain
\begin{equation}
\frac{\partial^2{\bar f}}{\partial \gamma^2} = 
\frac{3T^*}{2\gamma}\left(\gamma^{-1}-4\pi^2\right),
\label{second_gamma:eq}
\end{equation}
which is positive for $\gamma = \gamma_1$ but negative for
$\gamma = \gamma_2$. Only the first solution corresponds to 
a minimum and thus to freezing,
whereas the second is a saddle point. 
Within the limits of the first-RLV-shell approximation,
the crystals formed by $Q^{\pm}$-potentials feature thus a
{\it universal localization parameter} at freezing: irrespective
of the location on the freezing line and 
even {\it of the interaction potential itself}, the localization
length $\ell$ at freezing is a fixed fraction of the lattice constant
and the parameter $\gamma = (\alpha a^2)^{-1}$ attains along the
entire crystallization line the value
\begin{equation}
\gamma_{\rm f} \cong 0.018.
\label{gammaf:eq}
\end{equation}

We can
understand the physics behind the 
constancy of the ratio $\ell/a$ by examining anew the 
variational form of the free energy, Eq.\ (\ref{fren_var:eq}).
Suppose we have a fixed density $\rho^*$ and we vary $n_c$, seeking
to achieve a minimum of $\tilde f$. An increase in $n_c$ implies
an increase in the lattice constant $a$ by virtue of Eq.\ (\ref{den_a:eq}).
The density profile takes advantage of the additional space 
created between neighboring sites and becomes more delocalized. This
increase of the spreading of the profile brings with it an
entropic gain which exactly compensates the corresponding loss
from the accumulation of particles on a single site, expressed
by the term $\ln n_c$ in Eq.\ (\ref{fren_var:eq}). Expressing
$\ell$ in units of $a$, i.e., working with the variable $\gamma$
instead with the original one, $\alpha^*$, brings the additional
advantage that $\gamma_{\rm f}$ becomes independent of the 
pair potential. The corresponding value of $\alpha^*$ at freezing,
$\alpha^*_{\rm f}$, can be obtained from Eqs.\ (\ref{den_a:eq}) 
and (\ref{proport:eq}),
and reads for the bcc-lattice as
\begin{equation}
\alpha^*_{\rm f} = \frac{y_*^2}{8\pi^2\gamma_{\rm f}}.
\label{alphaf:eq}
\end{equation}
Here, a dependence on the pair interaction appears through 
the value of $y_*$.

Complementary to the localization parameter,
we can consider the Lindemann ratio $L$ at freezing,\cite{lindemann}
taking into account that for the bcc lattice the nearest neighbor
distance is $d = a\sqrt{3}/2$. Employing 
the Gaussian density parametrization,
we find $\langle r^2 \rangle = 3/(2\alpha)$ and thus 
$L \equiv \sqrt{\langle r^2 \rangle}/d  = \sqrt{2\gamma}$. 
Using (\ref{gammaf:eq}), the
Lindemann ratio at freezing, $L_{\rm f}$, is determined as
\begin{equation}
L_{\rm f} \cong 0.189.
\label{lind:eq}
\end{equation}
This value is considerably larger than the typical value of $0.10$ usually
quoted for systems with harshly repulsive particles, such as,
e.g., the bcc alcali metals and the fcc metals Al, Cu, Ag, and 
Au [\onlinecite{shapiro}], but close to the value 0.160 found
by Stillinger and Weber\cite{still3} for the Gaussian core model.
The particles in the cluster crystal are quite more delocalized
than the ones for singly-occupied solids. The clustering strategy
enhances the stability of the crystal with respect to oscillations
about the equilibrium lattice positions. 

The locus of freezing points $(T^*_{\rm f}, \rho^*_{\rm f})$
is easily obtained by Eqs.\ (\ref{dfzero:eq}) 
and (\ref{gamma:eq}) and
takes the form of of a straight line:
\begin{equation}
\frac{T^*_{\rm f}}{\rho^*_{\rm f}} = 16\pi^2\gamma_{\rm f}
                                |\tilde\phi(y_*)|e^{-4\pi^2\gamma_{\rm f}}
\cong  1.393\, |\tilde\phi(y_*)|. 
\label{freeze:eq}
\end{equation}
Contrary to the Lindemann ratio, which is independent of 
the pair potential, the freezing line does depend on the interaction
potential between the particles. Yet, this dependence is a 
particular one, as it rests exclusively on the absolute 
value of the Fourier transform at the minimum,
$|\tilde\phi(y_*)|$ and is simply proportional to it.
Comparing with the location of the $\lambda$-line from
Eq.\ (\ref{lambda:eq}),
$T^*_{\lambda}/\rho^*_{\lambda} = |\tilde\phi(y_*)|$, 
we find that crystallization
{\it preempts} the occurrence
of the instability: indeed, at fixed $T^*$, $\rho^*_{\rm f}
< \rho^*_{\lambda}$ or, equivalently, at fixed $\rho^*$,
$T^*_{\rm f} > T^*_{\lambda}$; see also Fig.\ \ref{phdg:fig}.
The transition is first-order, as witnessed by the jumps
of the values of $\alpha$ and $\rho_{\bf K}$ at the
transition, which
are nonzero for the crystal but vanish in the fluid.
This is analogous to the conjectured preemption of the Kirkwood instability
for infinite-dimensional hard spheres by a first-order freezing
transition.\cite{frisch:pre:99}

The freezing properties of $Q^{\pm}$-potentials are, thus,
quite unusual and at the same time quite simple: the lattice
constant is fixed due to a clustering mechanism that drives
the aggregation number $n_c$ proportional to the density.
The constant of proportionality depends solely on the wavenumber
$y_*$ for which the Fourier transform of the pair interaction
has a negative minimum, Eq.\ (\ref{proport:eq}).
The freezing line is a straight line
whose slope depends only on the value of the Fourier transform
of the potential at the minimum, Eq.\ (\ref{freeze:eq}). The Lindemann ratio
at freezing is a universal number, independent of interaction
potential and thermodynamic state. 

Whereas the Lindemann ratio is employed as a measure of the
propensity of a crystal to melt, the height of the peak of the structure
factor of the fluid is looked upon as a measure of the tendency
of the fluid to crystallize. The Hansen-Verlet 
criterion\cite{hv1,hv2}
states
that crystallization takes place when this quantity exceeds
the value 2.85. For the systems at hand, the maximum of $S(y)$
lies at $y_*$, as is clear from Eq.\ (\ref{sofk:eq}). 
Using Eq.\ (\ref{freeze:eq})
for the location of the freezing line, we obtain the value
$S_{\rm f}(y_*)$ on the freezing line as
\begin{equation}
S_{\rm f}(y_*) = \left[1 + \frac{\tilde\phi(y_*)}{1.393\,|\tilde\phi(y_*)|}
\right]^{-1} \cong 3.542.
\end{equation}
This value is considerably larger than the Hansen-Verlet threshold.\cite{hv1,hv2}
In the fluid phase,
$Q^{\pm}$-systems can therefore sustain a higher
degree of spatial correlation before they crystallize
than particles with diverging interactions do. This 
property lies in the fact that some contribution to
the peak height comes from correlations from {\it within} the clusters
that form in the fluid; the formation of clusters already in 
the uniform phase is witnessed by the maxima of $g(r)$ at $r=0$
seen in Fig.\ \ref{gofr:fig} and also explicitly
visualized in our previous simulations of the model.\cite{bianca:prl:06} 
These, however,
do not contribute to intercluster ordering that leads to 
crystallization. 
At any rate, 
the Hansen-Verlet peak height is also a 
universal quantity for all $Q^{\pm}$ systems, in the
framework of the current approximation. Moreover, both for the Lindemann
and for the Hansen-Verlet 
criteria, the $Q^{\pm}$ systems are more robust than usual ones,
since they allow for stable fluids with peak heights that exceed
$2.85$ by 25\% 
and for stable crystals with Lindemann ratios that
exceed $0.10$ by almost 90\%.

\section{Comparison with numerical minimization}
\label{compare:sec}

The density functional of Eq.\ (\ref{totalfren:eq}) is very accurate
for the bounded ultrasoft potentials 
at hand.\cite{criterion,likos:gauss,louis:mfa,archer1,archer2,archer3,archer4,finken} The modeling of
the inhomogeneous density as a sum of Gaussians is
an approximation but, again, an accurate one, as has been shown
by comparing with simulation results,\cite{bianca:prl:06}
see also section
\ref{harmonic:sec} of this work. The analytical results
derived in the preceding section rest on one additional approximation,
namely on ignoring the RLVs beyond the first shell. Here, we
want to compare with a full minimization of the functional
(\ref{totalfren:eq}) under the modeling of the density via
(\ref{gaussian_real:eq}), so as to test the accuracy of the
hitherto drawn conclusions on clustering and crystallization.

\begin{figure}
 \begin{center}
 \includegraphics[width=8.5cm, clip]
        {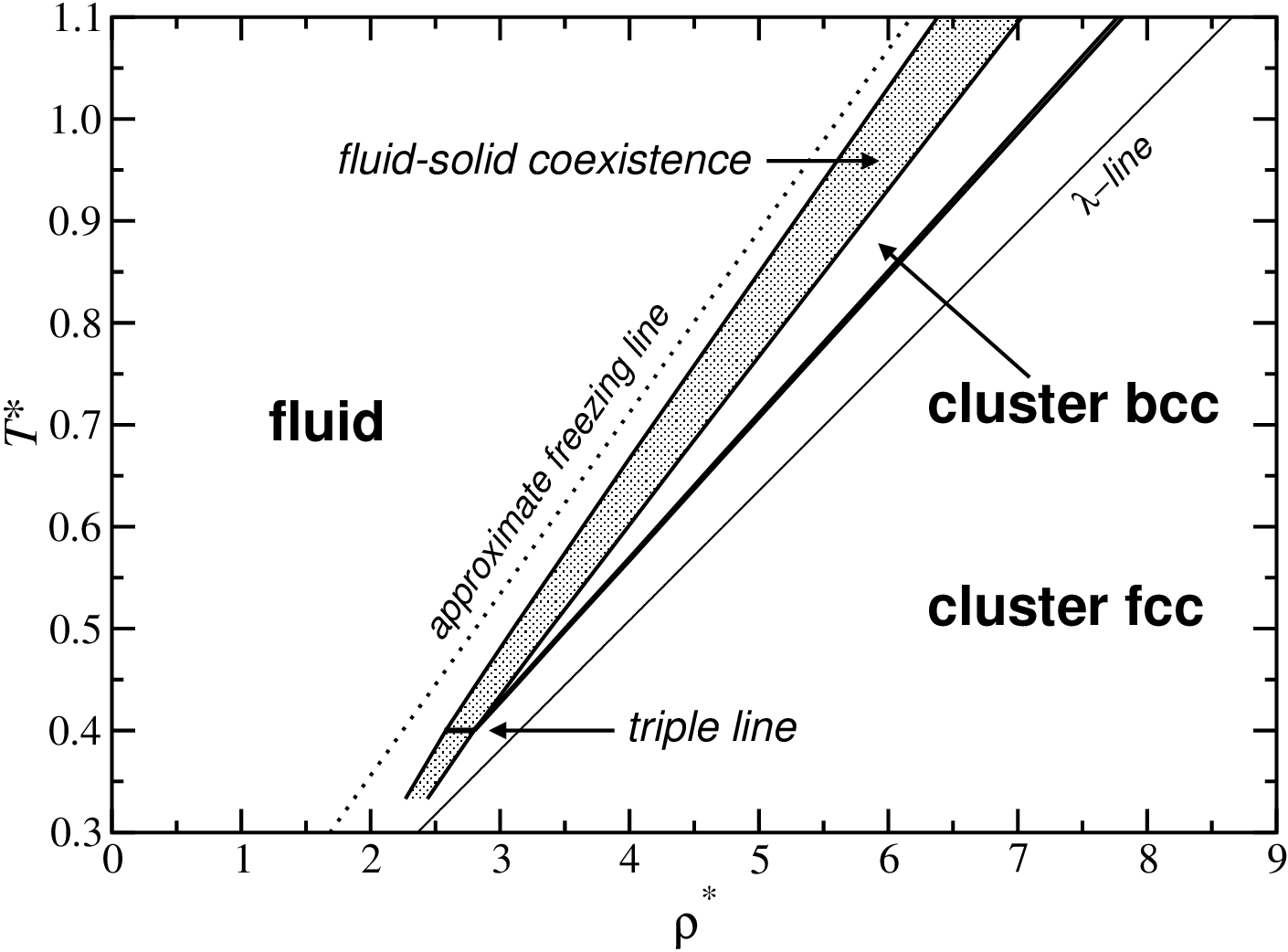}
 \end{center}
\caption{The phase diagram of the GEM-4 model, obtained by full
minimization of the density functional (\ref{totalfren:eq}), under
the Gaussian parametrization of the density, Eq.\ (\ref{gaussian_real:eq}),
redrawn from Ref.\ [\onlinecite{bianca:prl:06}]. On the same plot, we show by the
dotted line the 
approximate analytical result
for the freezing line, Eq.\ (\ref{freeze:eq}), as well as the
$\lambda$-line of the system, Eq.\ (\ref{lambda:eq}).}
\label{phdg:fig}
\end{figure}

We work with the concrete GEM-4 system, for which the minimization
of the density functional has been carried out
and the phase diagram has been calculated in Ref.\ [\onlinecite{bianca:prl:06}].
In Fig.\ \ref{phdg:fig} we show the phase diagram obtained by the
full minimization, compared with the freezing line from the 
analytical approximation, Eq.\ (\ref{freeze:eq}), for this system.
It can be seen that the latter is a very good approximation
to the full result, its quality improving 
slowly as the temperature grows; the analytical approximation
consistently overestimates the region of stability of the crystal.
Moreover, whereas the approximation only predicts a stable bcc
crystal, the high-density phase of the system is fcc. Although
bcc indeed is, above the triple temperature, the stable crystal
immediately post-freezing, it is succeeded at higher densities by
a fcc lattice, which our analytical theory fails to predict. 

\begin{figure}
 \begin{center}
 \includegraphics[width=8.5cm, clip]
        {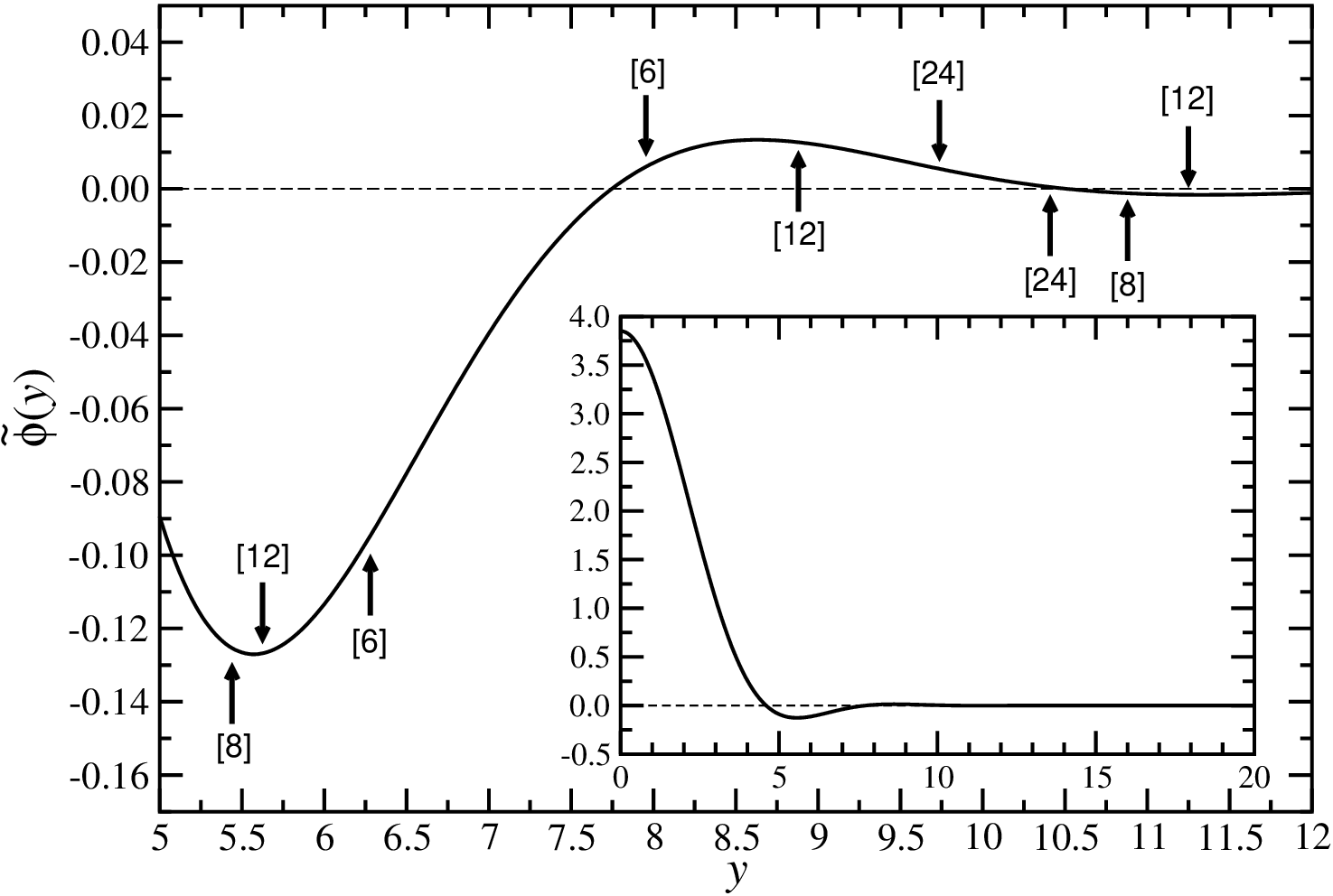}
 \end{center}
\caption{Inset: the Fourier transform $\tilde\phi(y)$ of the GEM-4
potential. Main plot: a zoom at the region of $\tilde\phi(y)$ in which
the first few nonvanishing RLV shells of the cluster-crystals of Fig.\ \ref{phdg:fig}
lie. The arrows denote the positions of the shells and the numbers
in square brackets the numbers of distinct RLVs within each shell.
These positions are the result of the full minimization of the
density functional. Downwards pointing arrows pertain to the direct bcc
lattice and upwards pointing arrows to the direct fcc one. In agreement
with clustering predictions, the positions of the RLVs are
density-independent.}
\label{phiofq:fig}
\end{figure}

All these discrepancies can be easily understood by looking at the
effects of ignoring the higher RLV shells from the summation
in the excess free energy, Eq.\ (\ref{fren_var:eq}). 
Consider first exclusively the bcc
lattice. In Fig.\ \ref{phiofq:fig} we show the locations of 
the bcc-RLVs, as obtained from the full minimization, by the
downwards pointing arrows. It can be seen that the first shell
is indeed located very closely to $y_*$, as the analytical 
solution predicts. However, the next two RLV shells do have
contributions and, due to their location on the hump of
$\tilde\phi(y)$, the latter is positive. By ignoring them
in performing the analytical solution, we are artificially lowering the
free energy of the crystal, increasing thereby its domain
of stability. 

The occurrence of a fcc-lattice that beats the bcc at high 
densities is only slightly more complicated to understand. 
A first remark is that the parameter $\alpha^*$ 
increases proportionally to $\rho^*/T^*$, see the following
section. Hence, the Gaussian factors 
from RLVs beyond the first shell, $\exp[-Y_i^2/(2\alpha^*)]$, $i \geq 2$,
gain weight in the sum as density grows. The cutoff for the RLV-sum is
now provided rather by the short-range nature of 
$\tilde\phi(y)$ than by the exponential factors.
Due to the increased importance of the contributions from the $i \geq 2$-terms
in the excess free energy sum, the relative location of higher RLVs
becomes crucial and can
tip the balance in favor of fcc, although the bcc-lattice
has a {\it higher} number of RLVs in its first shell than the fcc.
In Fig.\ \ref{phiofq:fig} we see that this is precisely what
happens: the second RLV shell of the fcc is located fairly
close to the first. In the full minimization, both of them
arrange their positions so as to lie close enough to 
$y_*$. Now, a total of fourteen 1st- and 2nd-shell RLVs
of the fcc can beat the twelve 1st-shell RLVs of the bcc
and bring about a structural phase transformation from the 
latter to the former. 

The relative importance of the first and second
neighbors is quantified by the ratio
\begin{equation}
\Delta = \frac{\xi_2}{\xi_1}\frac{|\tilde\phi(Y_2)|}{|\tilde\phi(Y_1)|}
\exp[-(Y_2^2-Y_1^2)/(2\alpha^*)],
\label{ratio:eq}
\end{equation}
where $\xi_2$ is the number of RLVs in the second shell.
If $\tilde\phi (y)$
does not decay sufficiently fast to zero as $y$ grows,
then the fcc lattice might even win over the bcc everywhere,
since then $\Delta$ could be considerable even for values of
$\alpha^*$ close to freezing, which are not terribly high.
In fact, the penetrable sphere model (GEM-$m$ with $m \to \infty$)
does not possess, on these grounds, 
a stable bcc phase at all.\cite{pensph:98}
The prediction of the analytical theory
on bcc stability has to be taken with care and is conditional
to $\Delta$ being sufficiently small. A quantitative criterion
on the smallness of $\Delta$ is model specific and cannot be
given in general. 
The determination of the stable phases of the GEM-$m$ family
and their dependence on $m$ can be achieved by employing
genetic algorithms\cite{dieter:jcp:05} and will be presented
elsewhere.\cite{bianca:long}

Notwithstanding the quantitative discrepancies between
the simplified, analytically tractable version of DFT and the
full one, which are small in the first place,
the central conclusion of the former remains intact:
the RLVs of the crystals are density-independent. Whereas 
the analytical approximation predicts that the 
length of the first RLV shell 
coincides with $y_*$, the numerical minimization brings about
small deviations from this prediction. However, by reading off
the relevant values from Fig.\ \ref{phiofq:fig}, we obtain
$Y_1 = 5.625$ for the bcc and $Y_1 = 5.441$ for the fcc-lattice
of the GEM-4 model.
Comparing with the ideal value
$y_* = 5.573$, we find that the deviation between them is only 
a few percent. Clustering takes place, so that the lattice
constants of both lattices remain fixed, a characteristic 
that was also explicitly confirmed by computer simulations
of the model.\cite{bianca:prl:06}

\section{Connection to harmonic theory of crystals}
\label{harmonic:sec}

The use of a Gaussian parametrization for the 
one-particle density profiles, Eq.\ (\ref{gaussian_real:eq}),
is a standard modeling of the latter for periodic crystals.
This functional form is closely related to the harmonic theory of
crystals.\cite{ashcroft} Each particle performs 
oscillations around its lattice site, experiencing thereby
an effective, one-particle site potential, $V_{\rm site}({\bf s})$
that is quadratic in the displacement $s$, for small values
$s/a$ [\onlinecite{foot2}].
Here, we will explicitly demonstrate that the Gaussian form
with the localization parameter predicted from density functional
theory coincides with the results obtained by performing 
a harmonic expansion of the said site potential.

The formation of clustered crystals is a generic property
of all $Q^{\pm}$-systems, since the $\lambda$-instability
is common to all of them; the form of the clusters that
occupy the lattice sites, however, can be quite complex,
depending on the details of the interaction. The Gaussian
parametrization (\ref{gaussian_real:eq}) implies that for
each of the $n_c$ particles of the cluster, the lattice 
site ${\bf R}$ is an equilibrium position. In other words,
the particular clusters we consider here are internally
structureless. Clusters with a well-defined internal order
have been found when an additional hard core of small extent is
introduced.\cite{primoz:07}  
A necessary 
requirement for the lack of internal order is that the Laplacian
of the interaction potential $v(r)$ be finite at 
$r=0$, as will be shown shortly.
On these grounds, we impose 
from the outset on the interaction potential
the additional requirement:
\begin{equation}
\nabla^2 v(r) = \frac{1}{r^2}\left(r^2 v'(r)\right)' 
< \infty\;\;{\rm for}\;\; r \to 0,
\label{condition1:eq}
\end{equation} 
where the primes denote the derivative with respect to $r$. 
Eq.\ (\ref{condition1:eq}) implies that $v'(r)$ must be
{\it at least} linear in $r$ as $r \to 0$. Concomitantly,
$v''(r)$ must be at least $O(1)$ as $r \to 0$. As a
consequence, we have
\begin{equation}
v''(r) - \frac{v'(r)}{r} \to 0\;\;{\rm for}\;\; r \to 0.
\label{condition:eq}
\end{equation}
It can be easily checked that (\ref{condition:eq}) is satisfied by
all members of the GEM-$m$ class for $m > 2$. It is also satisfied
by the $m=2$ member, i.e., the Gaussian model, which does
{\it not} display clustering because it belongs to the $Q^{+}$-class.
This is, however, no contradiction. As mentioned above, 
the condition (\ref{condition:eq}) is {\it necessary} for the
formation of structureless clusters and not a sufficient one.
For $Q^{\pm}$-potentials for which (\ref{condition:eq}) is not
fulfilled (such as the Fermi distribution models of 
Ref.\ [\onlinecite{criterion}]), this does not mean that 
clusters do not form; it rather points to the fact that
they possess some degree of internal order.

The clustered crystals can be considered as Bravais lattices with
a $n_c$-point basis. Accordingly, their phonon spectrum will 
feature 3 acoustic modes and $3(n_c - 1)$ optical modes, for which
the oscillation frequency $\omega(k)$ remains finite as $k \to 0$.
We are interested in the case $n_c \gg 1$, i.e., deep in the
region of stability of the crystal, where the clusters have 
a very high occupation number. Consequently, the phonon spectra
and the particle displacements will be dominated by the optical
branches. Further, we simplify the problem by choosing, in the
spirit of the Einstein model of the crystal,\cite{ashcroft} one 
specific optical phonon with $k = 0$
as a representative for the whole spectrum. This mode corresponds
to the relative partial displacement of two sublattices:
one with $n_c - 1$ particles on each site and one with just
the remaining one particle per site. The two sublattices
coincide at the equilibrium position and maintain their
shape throughout the oscillation mode, consistent with the
fact of an infinite-wavelength mode, $k=0$. Accordingly, the
site potential felt by any one of the particles of the 
single-occupied sublattice, $V_{\rm site}({\bf s})$, can
be expressed as
\begin{equation}
V_{\rm site}({\bf s}) = (n_c - 1)\left[v(s) + \sum_{{\bf R} \ne 0}
v(|{\bf s} - {\bf R}|)
\right],
\label{vsite:eq}
\end{equation}
where ${\bf s}$ is the relative displacement of the two sublattices.
For brevity, we also define 
\begin{equation}
W_{\rm site}({\bf s}) \equiv \frac{V_{\rm site}({\bf s})}{n_c-1}.
\label{wsite:eq}
\end{equation}

The Taylor expansion of a scalar function $f$
around a reference point ${\bf r}$ reads as
\begin{equation}
f({\bf r} + {\bf s}) = f({\bf r}) +
{\bf s}\cdot\nabla f({\bf r}) +
\frac{1}{2}\left({\bf s}\cdot\nabla\right)^2 f({\bf r}) + \cdots 
\label{taylor3:eq}
\end{equation}  
Setting ${\bf r} \to 0$ and $f \to W_{\rm site}$, we obtain the
quadratic expansion of the site potential; the constant $V_{\rm site}(0)$
is unimportant. For the linear term, we have
\begin{equation}
\nabla W_{\rm site}({\bf r}=0) = v'(r){\bf {\hat r}}\Big|_{{\bf r} = 0} 
+ \sum_{{\bf R} \ne 0} v'(R){\bf {\hat R}},
\label{linear:eq}
\end{equation}
where ${\bf {\hat r}}$ and ${\bf {\hat R}}$ are unit vectors. 
The sum in (\ref{linear:eq}) vanishes due to lattice inversion
symmetry; the first term also, since $v'(0) = 0$. Thus, the term
linear in ${\bf s}$ in the expansion of $V_{\rm site}({\bf s})$
vanishes, consistently with the fact that ${\bf s} = 0$ is 
an equilibrium position. 

We now introduce
Cartesian coordinates and write 
${\bf r} = (x,y,z)$, ${\bf s} = (s_x,s_y,s_z)$, and
${\bf R} = (R_x, R_y, R_z)$. The quadratic term in 
(\ref{taylor3:eq}) takes the explicit form
\begin{eqnarray}
\nonumber
\frac{1}{2}\left({\bf s}\cdot\nabla\right)^2 & = &
      s_xs_y\frac{\partial^2}{\partial x\partial y} +
      s_ys_z\frac{\partial^2}{\partial y\partial z} +
      s_zs_x\frac{\partial^2}{\partial z\partial x}
\\
& + &
\frac{1}{2}\left(s_x^2\frac{\partial^2}{\partial x^2} +
                 s_y^2\frac{\partial^2}{\partial y^2} +
                 s_z^2\frac{\partial^2}{\partial z^2}\right).
\label{explicit:eq}
\end{eqnarray}
Let us consider first the mixed derivative acting on $W_{\rm site}({\bf r})$,
evaluated at ${\bf r} = 0$. Using Eqs.\ (\ref{wsite:eq}) 
and (\ref{vsite:eq}), we obtain
\begin{eqnarray}
\nonumber
\frac{\partial^2 W_{\rm site}({\bf r})}{\partial x\partial y}\Big|_{{\bf r} = 0}
& = &  
\left[\frac{xy}{r^2}\left(v''(r)-\frac{v'(r)}{r}\right)\right]_{{\bf r} = 0}
\\
& + & \sum_{{\bf R}\ne 0}\frac{R_xR_y}{R^2}
\left[v''(R)-\frac{v'(R)}{R}\right].
\label{mixed:eq}
\end{eqnarray}
The first term on the right-hand side of (\ref{mixed:eq}) vanishes by
virtue of (\ref{condition:eq}). The second one also vanishes due to
the cubic symmetry of the lattice. Clearly, the other two terms in
(\ref{explicit:eq}) 
with mixed derivatives vanish as well. For the remaining terms,
the cubic symmetry of
the lattice implies that all three 
second partial derivatives of $W_{\rm site}$ at ${\bf r} = 0$
are equal:
\begin{eqnarray}
\nonumber
\frac{\partial^2 W_{\rm site}({\bf r})}{\partial x^2}\Big|_{{\bf r} = 0}
& = & 
\frac{\partial^2 W_{\rm site}({\bf r})}{\partial y^2}\Big|_{{\bf r} = 0}
=
\frac{\partial^2 W_{\rm site}({\bf r})}{\partial z^2}\Big|_{{\bf r} = 0}
\\
& = &  \frac{1}{3}\nabla^2W_{\rm site}({\bf r})\Big|_{{\bf r} = 0}.
\label{laplace:eq}
\end{eqnarray}

Gathering the results, we obtain the expansion of $V_{\rm site}({\bf s})$
to quadratic order in $s$ as
\begin{equation}
V_{\rm site}({\bf s}) = V_{\rm site}(0) + \left[\frac{(n_c - 1)}{6}
\sum_{\bf R}\nabla^2v(R)\right]s^2,
\label{isotropic:eq}
\end{equation}
which is isotropic in $s$, as should for a crystal of cubic symmetry.
The one-particle motion is therefore harmonic; as we consider
$n_c \gg 1$, we set $n_c - 1 \cong n_c$ in (\ref{isotropic:eq})
and we obtain the effective, one-particle Hamiltonian 
${\mathcal H}_1$ in the form:
\begin{equation}
{\mathcal H}_1 = \frac{p^2}{2m} + \kappa s^2,
\label{hamilton:eq}
\end{equation}
with
\begin{equation}
\kappa = \frac{n_c}{6}\sum_{\bf R}\nabla^2 v(R)
\label{kappa:eq}
\end{equation}
and the momentum $p$ and mass $m$ of the particle. The density
profile $\rho_1({\bf r})$ of this single-particle problem is
easily calculated as 
$\langle \delta({\bf r} - {\bf s})\rangle_{{\mathcal H}_1}$, yielding
\begin{equation}
\rho_1({\bf r}) = \left(\frac{\beta\kappa}{\pi}\right)^{3/2}
e^{-\beta\kappa r^2}.
\label{rho1:eq}
\end{equation}
This is indeed a Gaussian of a single
particle, with a localization parameter
$\alpha_{\rm h} = \beta\kappa$; the total density on a given
site will be then just $n_c\rho_1({\bf r})$, in agreement with the
functional form put forward in Eq.\ (\ref{gaussian_real:eq}).

It is useful to consider in detail the form of the
localization parameter $\alpha_{\rm h}$ predicted by the 
harmonic theory. The parameter $\kappa$ is expressed as
a sum of the values of $\psi(r) = \nabla^2 v(r)$ over 
the periodic set $\{{\bf R}\}$. For every function 
$\psi(r)$ that possesses a Fourier transform
$\tilde\psi(k)$, it holds\cite{ashcroft}
\begin{equation}
\sum_{{\bf R}} \psi(R) = \rho_s\sum_{{\bf K}} 
\tilde\psi(K),
\label{parseval:eq}
\end{equation}
where $\{{\bf K}\}$ is the set of RLVs of $\{{\bf R}\}$ and
$\rho_s = N_s/V$ is the density of {\it lattice sites} of $\{{\bf R}\}$.
From (\ref{nc:eq}),
$\rho_s n_c = \rho$. 
Taking into account that the Fourier transform of $\nabla^2 v(r)$
is $-k^2\tilde v(k)$, we obtain for the localization parameter
of the harmonic theory the result
\begin{equation}
\alpha_{\rm h} = -\frac{\rho}{6k_{\rm B}T}\sum_{{\bf K}}K^2\tilde v(K).
\label{alphah:eq}
\end{equation}
The localization parameter must be, evidently, positive.
Eq.\ (\ref{alphah:eq}) manifests the impossibility for cluster
formation if the Fourier transform of the pair potential is
nonnegative, i.e., for $Q^{+}$ interactions. In the 
preceding section, we showed within the DFT formalism that if
the potential is $Q^{\pm}$, this implies the formation of
cluster crystals. Harmonic theory allows us to make the
opposite statement as well: if the potential is {\it not}
$Q^{\pm}$, then there can be {\it no clustered crystals}.
Therefore, an {\it equivalence} between the $Q^{\pm}$ 
character of the interaction and the formation of clustered
crystals can be established. Moreover, Eq.\ (\ref{alphah:eq})
offers an additional indication as to why the RLV where
$\tilde v(K)$ is most negative is selected as the shortest
nonvanishing one by the clustered
crystals: this is the best strategy in order to keep the localization
parameter positive. 

Harmonic theory provides, therefore, 
an insight into the necessity of locating the first shell 
of the RLVs at $k_*$ from a different point of view than
density functional theory does.
The choice $K_1 = k_*$ guarantees that the particles
inhabiting neighboring clusters provide the restoring forces that
push any given particle back towards its equilibrium position.
The density functional treatment of the preceding sections
establishes that 
the lattice constant is chosen by $Q^{\pm}$-systems in such 
a way that the sum of {\it intracluster} and
{\it intercluster} interactions, together with the entropic
penalty for the aggregation of $n_c$ particles is optimized.\cite{bianca:long}
The unlimited growth of $n_c$ is avoided by the requirement of
mechanical stability of the crystal. Indeed, for too high 
$n_c$-values, the lattice constant would concomitantly grow,
so that the resulting
restoring forces working against the thermal fluctuations,
would become too weak to sustain the particles at their
equilibrium positions.

Let us, finally, compare the result (\ref{alphah:eq}) for
the localization parameter with the prediction from DFT.
We consider the high-density crystal phase, for which
$\alpha^*$ is very large, so that the simplification that only
the first shell of RLVs can be kept in (\ref{fren_var:eq}) must be dropped.
Setting $\partial\tilde f/\partial\alpha^* = 0$ there, we
obtain
\begin{equation}
\frac{3T^*}{2\alpha^*} + \frac{\rho^*}{\left(2\alpha^*\right)^2}
\sum_{{\bf Y}}Y^2\tilde\phi(Y)e^{-Y^2/(2\alpha^*)} = 0.
\label{dftall:eq}
\end{equation}
The function $\tilde\phi(y)$ is short-ranged in
reciprocal space, thus the sum in (\ref{dftall:eq}) 
can be effectively truncated at some finite upper cutoff
$Y_c$. Then, there exists a sufficiently large 
density $\rho^*$ beyond which the 
parameter $\alpha^*$ is so large that
$Y^2/(2\alpha^*) \ll 1$ for all $Y \leq Y_c$ included in
the summation. Accordingly, we can approximate all exponential factors
with uniti in (\ref{dftall:eq}), obtaining an algebraic equation
for $\alpha^*$. Reverting back to dimensional quantities, its
solution reads as 
\begin{equation}
\alpha = -\frac{\rho}{6k_{\rm B}T}\sum_{{\bf K}}
K^2\tilde v(K),
\label{alphasdft:eq}
\end{equation}
and is {\it identical} with the result from the harmonic theory,
Eq.\ (\ref{alphah:eq}). Thus, density functional theory and
harmonic theory become identical to each other at the limit of
high localization. This finding completes and generalizes the result of
Archer,\cite{archer} who established a close relationship between
the mean-field DFT and the Einstein model for the form of the
variational free energy functional of the system.

Consistently with our assumptions, 
$\alpha$ indeed grows with density. In fact, since the set of
RLVs in the sum of (\ref{alphasdft:eq}) is fixed, it can be
seen that $\alpha$ is simply proportional to $\rho/T$. 
This peculiar feature of 
the class of systems we consider 
is not limited to the localization parameter, and its significance
is discussed in the following section. 

\section{Connection with inverse-power potentials}
\label{invpower:sec}

As can be easily confirmed by the form of the density functional
of Eq.\ (\ref{totalfren:eq}),
the mean-field nature of the class of ultrasoft systems considered
here (both $Q^{+}$- and $Q^{\pm}$-potentials) implies that 
the structure and thermodynamics of the systems is fully determined
by the ratio $\rho^{*}/T^{*}$ between density and temperature and
not separately by $\rho^{*}$ and $T^{*}$. This is a particular
type of scaling between the two relevant thermodynamic variables,
reminiscent of the situation for systems interacting by means of
inverse-power-law potentials $v(r)$ having the form:
\begin{equation}
v(r) = \epsilon\left(\frac{\sigma}{r}\right)^n.
\label{powerlaw:eq}
\end{equation}
For such systems, it can be shown that their
statistical mechanics is governed by a 
a single coupling constant $\Gamma_D(n)$ expressed as\cite{weeks:prb:81}
\begin{equation}
\Gamma_D(n) = 
\frac{\rho^{*}}{\left({T^{*}}\right)^{D/n}},
\label{gamman:eq}
\end{equation}
where $D$ is the space dimension. It would appear that inverse-powers
$n=D$ satisfy precisely the same scaling as mean-field systems do
but there is a condition to be 
fulfilled: inverse-power systems are stable against
explosion, {\it provided} that $n > D$; this can be most easily seen
by considering the expression for the excess internal energy
per particle, $u(\rho,T)$, given by\cite{hansen:book}
\begin{equation}
u(\rho,T) = \frac{2\pi^{D/2}\rho\epsilon\sigma^n}{[(D/2)-1]!}
\int_0^{\infty}r^{D-1-n}g(r;\rho,T){\rm d}r.
\label{inten:eq}
\end{equation}
As $g(r) \to 1$ for $r \to \infty$, we see that the integral 
in (\ref{inten:eq}) converges only if $n > D$; a logarithmic
divergence results for $n=D$. A `uniform neutralizing background'
has to be formally introduced for $n \leq D$,
to obtain stable 
pseudo one-component systems, such as 
the one-component plasma.\cite{ocp1,ocp2,lieb}
Since we aim at staying with genuine one-component systems
throughout, we must strictly maintain $n > D$.  

Instead of taking the limit $n \to D$, we consider therefore a 
different procedure by setting
\begin{equation}
n = D + \delta 
\end{equation}
with some arbitrary, finite $\delta > 0$. Then the coupling constant
$\Gamma_D(n)$ becomes
\begin{equation}
\Gamma_D(D+\delta) = \rho^*\left(T^{*}\right)^{-\left(1+\delta/D\right)^{-1}}.
\label{gammam:eq}
\end{equation}  
Now take $D \to \infty$ in this prescribed fashion, obtaining:
\begin{equation}
\lim_{D\to\infty}\Gamma_D(D+\delta) = \frac{{\rho^{*}}}{{T^{*}}},
\label{gammainf:eq}
\end{equation}
which has precisely the same form as the coupling constant of 
our systems. In taking the 
limit $D \to \infty$, the exponent $n = D + \delta$ of the
inverse-power potential diverges as well. It can be easily
seen that in this case, the interaction $v(r)$ of 
Eq.\ (\ref{powerlaw:eq}) becomes a hard-sphere potential of
diameter $\sigma$. In other words, the procedure
prescribed above brings us once more to infinite-dimensional
hard spheres, because also the inverse-power interaction becomes
arbitrarily steep.
This is a very different way of taking the limit
than in Refs.\ [\onlinecite{frisch:pra:87,frisch:pre:99,bagchi:jcp:88}]: 
there, the interaction is hard 
in the first place and subsequently the limit
$D \to \infty$ is taken, 
whereas here, interaction and dimension of space change
together, in a well-prescribed fashion. 

The fact that the
statistical mechanics of ultrasoft fluids in three dimensions
is determined by the same dimensionless parameter as that
of a particular realization of hard spheres in infinite 
dimensions is intriguing. In a sense, ultrasoft systems
are effectively high-dimensional, since they allow for
extremely high densities, for which every particle interacts
with an exceedingly high number of neighbors. They might,
in this sense, provide for three-dimensional approximate
realizations of infinite-dimensional models. 
This is yet another relation to infinite-dimensional
systems, in addition to the one discussed at the end 
of Sec.\ \ref{dft:sec}.
Whether there
exists a deeper mathematical connection between the two classes,
remains a problem for the future. 

\section{Summary and concluding remarks}
\label{summary:sec}

We have provided a detailed analysis of the properties of 
bounded, ultrasoft systems, with emphasis on the $Q^{\pm}$-class
of interaction potentials. After having demonstrated the 
suppression of the contributions from the high-order direct correlation
functions of the fluid phases (of order 3 and higher), we
established as a consequence the accurate mean-field density
functional for arbitrary inhomogeneous phases. Though this
functional has been introduced and successfully used
in the recent past both in 
statics\cite{criterion,bianca:prl:06,likos:gauss,louis:mfa,archer1,archer2,
archer3,archer4,finken} 
and in dynamics,\cite{rex:molphys:06}
a sound justification of its basis on the properties of the
uniform phase was still lacking. 

The persistence of a {\it single}, finite length scale
for the lattice constants of the ensuing 
solids of $Q^{\pm}$-systems has been understood by a detailed
analysis of the structure of the free energy functional. 
In the fluid, the same length scale appears since the position
of maximum of the liquid structure factor is independent of 
density.
The negative minimum of the interaction potential in Fourier 
space sets this unique scale and forces in the crystal the
formation of clusters, whose population scales proportionally
with density. The analytical solution of an approximation of
the density functional is checked to be accurate when 
confronted with the full numerical minimization of the latter.
Universal Lindemann ratios and Hansen-Verlet values at crystallization
are predicted to hold for all these systems, which
differ substantially from those for hard matter systems.
The analytical 
derivation of these results provides useful insight into the
robustness of these structural values for an enormous variety
of interactions.

Though the assumption of bounded interactions has been made
throughout, recent results\cite{primoz:07}
indicate that both cluster formation and the persistence of
the length scale survive when a short-range diverging core is
superimposed on the ultrasoft potential, provided the range of
the hard core does not exceed, roughly, 20\% of the overall
interaction range.\cite{primoz:07} The morphology of the resulting
clusters is more complex, as full overlaps are explicitly
forbidden; even the macroscopic phases are affected, with
crystals, lamellae, inverted lamellae and `inverted crystals'
showing up at increasing densities. The generalization of our
density functional theory to such situations and the modeling
of the nontrivial, internal cluster morphology is a challenge
for the future. Here, a mixed density functional, employing
a hard-sphere and a mean-field part of the direct correlation
function seems to be a promising way to proceed.\cite{sear,lr1}
Finally, the study of the vitrification, dynamical arrest and
hopping processes in concentrated $Q^{\pm}$-systems is another
problem of current interest. The recent `computer synthesis' 
of model, amphiphilic dendrimers that do display precisely 
the form of $Q^{\pm}$-interactions discussed in this work,\cite{bianca:dendris}
offers concrete suggestions for the experimental realization
of the hitherto theoretically predicted phenomena.

\section*{ACKNOWLEDGMENTS}
We are grateful to Andrew Archer for helpful discussions and 
a critical reading of the manuscript and to Andras S{\"u}t{\H o}
for helpful comments.
This work has been supported by 
the {\"O}sterreichische Forschungsfond under
Project No.\ P17823-N08, as well as by the 
Deutsche Forschungsgemeinschaft within the
Collaborative Research Center SFB-TR6,
``Physics of Colloidal Dispersions in External Fields'', 
Project Section C3.

\appendix
\section{Proof that the GEM-$m$ models with $m > 2$ are 
$Q^{\pm}$-potentials}

Consider the inverse Fourier transform of the 
spherically symmetric, bounded pair
potential $v(r)$, reading as
\begin{equation}
v(r) = \frac{1}{2\pi^2}\int_0^{\infty}k^2\tilde v(k)\frac{\sin kr}{kr}{\rm d}k
\label{invft:eq}
\end{equation}
From (\ref{invft:eq}), it is straightforward to
show that the second derivative of $v(r)$ at $r=0$ takes
the form
\begin{equation}
v''(r=0) = -\frac{1}{6\pi^2}\int_0^{\infty}k^4\tilde v(k){\rm d}k.
\label{vpp:eq}
\end{equation}
Evidently, 
if $v''(r=0) \geq 0$, then $\tilde v(k)$ {\it must}
have negative parts and hence $v(r)$ is $Q^{\pm}$. For the 
GEM-$m$ family, it is easy to show that $v''(r=0) = 0$ for
$m > 2$, thus these members are indeed $Q^{\pm}$, as stated in the
main text. Double-Gaussian potentials of the form
\begin{equation}
v(r) = |\epsilon_1|e^{-(r/\sigma_1)^2} - |\epsilon_2|e^{-(r/\sigma_2)^2},
\label{dgauss:eq}
\end{equation}
with $|\epsilon_1| > |\epsilon_2|$, $\sigma_1 > \sigma_2$,
which feature a local {\it minimum}
at $r = 0$ are also $Q^{\pm}$, for the same reason. Notice,
however, that $v''(r=0) \geq 0$ is a sufficient, not a necessary
condition for membership in the $Q^{\pm}$-class. Thus, there exist
$Q^{\pm}$-potentials for which $v''(r=0) < 0$.

\section{Proof of the equivalence between the variational
free energies $\tilde f$ and $\bar f$.} 

The introduction of the new variable $\gamma$ instead of
$\alpha^*$, Eq.\ (\ref{amin_nc:eq}), and the subsequent new
form $\bar f$ of the variational free energy, Eq.\ (\ref{fbar:eq}),
are just a matter of convenience, which makes the minimization
procedure more transparent. A free gift of the
variable transformation is also the ensuing diagonal 
form of the Hessian matrix at the extremum. Fully equivalent
results are obtained, of course, by working with the original
variational free energy, $\tilde f$, Eq.\ (\ref{fren_var:eq}).
Here we explicitly demonstrate this equivalence.

Keeping, consistently, only the first shell of RLVs with length
$Y_1$, $\tilde f$ takes the form:
\begin{eqnarray}
\nonumber
\tilde f(n_c,\alpha^*;T^*,\rho^*)
& = &
T^*\left[\ln n_c + \frac{3}{2}\ln\left(\frac{\alpha^*}{\pi}\right)
- \frac{5}{2}\right] 
\\
\nonumber
& + &  \frac{\rho^*}{2}\tilde\phi(0) 
\\
& + &
 \frac{\xi_1\rho^*}{2}\tilde\phi(Y_1)
 e^{-Y_1^2/(2\alpha^*)},
\label{fren_new:eq}
\end{eqnarray}
where $Y_1$ and $n_c$ are related via Eq.\ (\ref{d:eq}). Minimizations
of $\tilde f$ with respect to $n_c$ and $\alpha^*$ yield,
respectively:
\begin{equation}
T^* -\frac{\xi_1\rho^*Y_1}{6}\left[\frac{\partial\tilde\phi(Y_1)}{\partial Y_1}
-\frac{Y_1}{\alpha^*}\tilde\phi(Y_1)\right]e^{-Y_1^2/(2\alpha^*)} = 0,
\label{delnc:eq}
\end{equation}
and
\begin{equation}
T^* + \frac{\xi_1\rho^*}{6\alpha^*}
\tilde\phi(Y_1)Y_1^2e^{-Y_1^2/(2\alpha^*)} = 0.
\label{dela:eq}
\end{equation}
Subtracting the last two equations from one another we obtain
\begin{equation}
\frac{\xi_1\rho^*}{6}\tilde\phi'(Y_1)Y_1
e^{-Y_1^2/(2\alpha^*)} = 0,
\end{equation}
implying $Y_1 = y_*$, as in the main text (once more, $Y_1 = 0$
is a formal solution that must
be rejected on the same grounds mentioned in the text.) From this property, 
Eq.\ (\ref{proport:eq})
immediately follows. Introducing 
$\Delta \tilde f \equiv \tilde f - f_{\rm liq}$, we can 
determine $\alpha^*_{\rm f}$
on the freezing line by requiring the simultaneous satisfaction of 
the minimization conditions above and of $\Delta\tilde f = 0$. The latter
equation yields
\begin{eqnarray}
\nonumber
T^*\left[\ln\left(\frac{n_c}{\rho^*}\right) 
+ \frac{3}{2}\left[\ln\left(\frac{\alpha^*_{\rm f}}{\pi}\right) - 1\right]\right]
\\
 =  - \frac{\xi_1\rho^*}{2}\tilde\phi(y_*)e^{-y_*^2/(2\alpha^*_{\rm f})},
\label{delftildezero:eq}
\end{eqnarray}
which, together with Eq.\ (\ref{dela:eq}), yields, after some
algebra
\begin{equation}
\ln\left[
\left(\frac{\alpha^*_{\rm f}}{\pi}\right)
\left(\frac{n_c}{\rho^*}\right)^{2/3}\right] - 1
= \frac{2\alpha^*_{\rm f}}{y_*^2}.
\label{onemore:eq}
\end{equation}
Using Eqs.\ (\ref{amin_nc:eq}) and (\ref{y1nc:eq}), we 
obtain the equation for the $\gamma_{\rm f}$-parameter at freezing as
\begin{equation}
-\left[
\ln\left(z^{2/3}\gamma_{\rm f}\pi \right) + 1\right] = 
\frac{2}{\gamma_{\rm f}\zeta^2},
\label{gamma_new:eq}
\end{equation}
which, upon setting $z=2$ and $\zeta = 2\sqrt{2}\pi$, yields
Eq.\ (\ref{gamma:eq}) of the main text. Alternatively, we can
introduce the 
variable $t \equiv \alpha^*_{\rm f}/y_*^2$ and rewrite Eq.\ (\ref{onemore:eq}) 
as
\begin{equation}
\ln\left[\left(8\sqrt{2}\right)^{2/3}\pi t\right] - 1 = 2t,
\label{omega:eq}
\end{equation}
which delivers $t \cong 0.704$ as a 
solution or, equivalently, $t = (8\pi^2\gamma_{\rm f})^{-1}$,
in agreement with Eqs.\ (\ref{gammaf:eq}) and (\ref{alphaf:eq}) of the main text.


\begin{thebibliography}{99}

\bibitem{sear} R. Sear and W. M. Gelbart, J. Chem. Phys. {\bf 110},
4582 (1999).

\bibitem{fs1} F. Sciortino, S. Mossa, E. Zaccarelli, and 
P. Tartaglia, Phys. Rev. Lett. {\bf 93}, 055701 (2004).

\bibitem{fs2} S. Mossa, F. Sciortino, P. Tartaglia,
and E. Zaccarelli, Langmuir {\bf 20}, 10756 (2004).

\bibitem{bartlett1} A. I. Campbell, V. J. Anderson, J. S. van Duijneveldt,
and P. Bartlett, Phys. Rev. Lett. {\bf 94}, 208301 (2005).

\bibitem{bartlett2} R. Sanchez and P. Bartlett, J. Phys.: Condensed Matter
{\bf 17}, S3351 (2005).

\bibitem{lr1} A. Imperio and L. Reatto, J. Phys.: Condens. Matter
{\bf 16}, S3769 (2004).

\bibitem{lr2} A. Imperio and L. Reatto, J. Chem. Phys. {\bf 124},
164712 (2006).

\bibitem{jpcm} C. N. Likos, C. Mayer, E. Stiakakis, and G. Petekidis,
J. Phys.: Condens. Matter {\bf 17}, S3363 (2005).

\bibitem{epl} E. Stiakakis, G. Petekidis, D. Vlassopoulos, 
C. N. Likos, H. Iatrou, N. Hadjichristidis, and J. Roovers, 
Europhys. Lett. {\bf 72}, 664 (2005).

\bibitem{likos:pr:01} C. N. Likos, Phys. Rep. {\bf 348}, 267 (2001).

\bibitem{krueger:89} B. Kr{\"u}ger, L. Sch{\"a}fer, and
A. Baumg{\"a}rtner, J. Phys. (France) {\bf 50}, 319 (1989).

\bibitem{hall:94} J. Dautenhahn and C. K. Hall, Macromolecules {\bf 27},
5399 (1994).

\bibitem{louis:prl:00} A. A. Louis, P. G. Bolhuis, J.-P. Hansen,
and E. J. Meijer, Phys. Rev. Lett. {\bf 85}, 2522 (2000).

\bibitem{ingo:jcp:04} I. O. G{\"o}tze, H. M. Harreis, and 
C. N. Likos, J. Chem. Phys. {\bf 120}, 7761 (2004).

\bibitem{ballik:ac:04} M. Ballauff and C. N. Likos, Angew. Chemie
Intl. English Ed. {\bf 43}, 2998 (2004).

\bibitem{denton:03} A. R. Denton, Phys. Rev. E {\bf 67}, 011804 (2003);
Erratum, ibid.\ {\bf 68}, 049904 (2003).

\bibitem{gottwald:04} D. Gottwald, C. N. Likos, G. Kahl, and H. L{\"o}wen,
Phys. Rev. Lett. {\bf 92}, 068301 (2004).

\bibitem{gottwald:05} D. Gottwald, C. N. Likos, G. Kahl, and H. L{\"o}wen,
J. Chem. Phys. {\bf 122}, 074903 (2005).

\bibitem{pierleoni:prl:06} C. Pierleoni, C. Addison, J.-P. Hansen,
and V. Krakoviack, Phys. Rev. Lett. {\bf 96}, 128302 (2006).

\bibitem{hansen:molphys:06} J.-P. Hansen and C. Pearson, Mol. Phys. {\bf 104},
3389 (2006).

\bibitem{suto:prl:05} A. S{\"u}t{\H o}, Phys. Rev. Lett. {\bf 95}, 265501 
(2005).

\bibitem{suto:prb:06} A. S{\"u}t{\H o}, Phys. Rev. B {\bf 74}, 104117 (2006).

\bibitem{klein:94} W. Klein, H. Gould, R. A. Ramos, I. Clejan,
and A. I. Melcuk, Physica (Amsterdam) {\bf 205A}, 738 (1994).

\bibitem{pensph:98} C. N. Likos, M. Watzlawek, and H. L{\"o}wen,
Phys. Rev. E {\bf 58}, 3135 (1998).

\bibitem{criterion} C. N. Likos, A. Lang, M. Watzlawek, and H. L{\"o}wen,
Phys. Rev. E {\bf 63}, 031206 (2001).

\bibitem{bianca:prl:06} B. M. Mladek, D. Gottwald, G. Kahl, M. Neumann,
and C. N. Likos, Phys. Rev. Lett. {\bf 96}, 045701 (2006); Erratum,
{\it ibid.}\ {\bf 97}, 019901 (2006).

\bibitem{stillinger:physica} F. H. Stillinger and D. K. Stillinger,
Physica (Amsterdam) {\bf 244A}, 358 (1997).

\bibitem{likos:gauss} A. Lang, C. N. Likos, M. Watzlawek, and H. L{\"o}wen,
J. Phys.: Condens. Matter {\bf 12}, 5087 (2000).

\bibitem{saija1} S. Prestipino, F. Saija, and P. V. Giaquinta,
Phys. Rev. E {\bf 71}, 050102 (2005),

\bibitem{saija2} F. Saija and P. V. Giaquinta, Chem. Phys. Chem. {\bf 6},
1768 (2005)

\bibitem{saija3} S. Prestipino, F. Saija, and P. V. Giaquinta,
J. Chem. Phys. {\bf 123}, 144110 (2005).

\bibitem{archer} A. J. Archer, Phys. Rev. E {\bf 72}, 051501 (2006).

\bibitem{daan:nature} D. Frenkel, Nature (London) {\bf 440}, 4-5 (02 March 2006).

\bibitem{daan:web} See the extensive commentary by D. Frenkel in:
Journal Club for Condensed Matter Physics, selection of March 2006,
link
http://www.bell-labs.com/jc-cond-mat/march/march\_2006.html.

\bibitem{primoz:07} M.~A.~Glaser, G.~M.~Grason, R.~D.~Kamien,
A.~Ko{\v s}mrlj, C.~D.~Santagelo, and P.~Ziherl, preprint,
cond-mat/0609570. 

\bibitem{Ruelle:book} D. Ruelle, {\it Statistical Mechanics}
(Benjamin, New York, 1969).

\bibitem{fernaud} M.-J. Fernaud, E. Lomba, and L. L. Lee,
J. Chem. Phys. {\bf 112}, 810 (2000).

\bibitem{hansen:book} J.-P. Hansen and I. R. McDonald,
{\it Theory of Simple Liquids}, 3rd ed. (Elsevier, Amsterdam, 2006).

\bibitem{percus:62} J. K. Percus, Phys. Rev. Lett. {\bf 8}, 462 (1962).

\bibitem{percus:64} J. K. Percus, in {\it The Equilibrium Theory of
Classical Fluids}, edited by H. L. Fritsch and J. L. Lebowitz
(Benjamin, New York, 1964).

\bibitem{denton:pra:91} A. R. Denton and N. W. Aschroft,
Phys. Rev. A {\bf 44}, 1219 (1991).

\bibitem{yethiraj:jcp:01} A. Yethiraj, H. Fynewever, and
C.-Y. Shew, J. Chem. Phys. {\bf 114}, 4323 (2001).

\bibitem{singh} Y. Singh, Phys. Rep. {\bf 207}, 351 (1991).

\bibitem{evans:78} R. Evans, Adv. Phys. {\bf 28}, 143 (1979).

\bibitem{ry:79} T. V. Ramakrishnan and M. Yussouff, Phys. Rev. B {\bf 19},
2775 (1979).

\bibitem{haymet:81} A. D. J. Haymet and D. W. Oxtoby, J. Chem. Phys. {\bf 74},
2559 (1981).

\bibitem{denton:pra:90} A. R. Denton and N. W. Ashcroft, Phys. Rev. A {\bf 41},
2224 (1990).

\bibitem{ocp1} M. Baus and J.-P. Hansen, Phys. Rep. {\bf 59}, 1 (1980).

\bibitem{ocp2} S. Ichimaru, H. Iyetomi, and S. Tanaka, 
Phys. Rep. {\bf 149}, 91 (1987).

\bibitem{barrat:prl:87} J.-L. Barrat, J.-P. Hansen, and G. Pastore,
Phys. Rev. Lett. {\bf 58}, 2075 (1987).

\bibitem{barrat:molphys:88} J.-L. Barrat, J.-P. Hansen, and G. Pastore,
Mol. Phys. {\bf 63}, 747 (1988).

\bibitem{denton:pra:89} A. R. Denton and N. W. Ashcroft, Phys. Rev. A {\bf 39},
426 (1989).

\bibitem{likos:pusey} C. N. Likos, N. Hoffmann, H. L{\"o}wen, and
A. A. Louis, J. Phys.: Condens. Matter {\bf 14}, 7681 (2002). 

\bibitem{ingo:06} I. O. G{\"o}tze, A. J. Archer, and C. N. Likos,
J. Chem. Phys. {\bf 124}, 084901 (2006).

\bibitem{louis:mfa} A. A. Louis, P. G. Bolhuis, and J.-P. Hansen,
Phys. Rev. E {\bf 62}, 7961 (2000).

\bibitem{archer1} A. J. Archer and R. Evans, Phys. Rev. E {\bf 64},
041501 (2001).

\bibitem{archer2} A. J. Archer and R. Evans, J. Phys.: Condens. Matter
{\bf 14}, 1131 (2002).

\bibitem{archer3} A. J. Archer, C. N. Likos, and R. Evans,
J. Phys.: Condens. Matter {\bf 14}, 12031 (2002).

\bibitem{archer4} A. J. Archer, C. N. Likos, and R. Evans,
J. Phys.: Condens. Matter {\bf 16}, L297 (2004).

\bibitem{finken} R. Finken, J.-P. Hansen, and A. A. Louis,
J. Phys. A {\bf 37}, 577 (2004). 

\bibitem{stell} G. Stell, J. Stat. Phys. {\bf 78}, 197 (1995).

\bibitem{ciach} A. Ciach, W. T. G{\' o}{\' z}d{\' z}, and R. Evans,
J. Chem. Phys. {\bf 118}, 9726 (2003).

\bibitem{kirkwood:51} J. G. Kirkwood, in {\it Phase Transitions in Solids},
ed. by R. Smoluchowski, J. E. Meyer, and A. Weyl (Wiley, New York, 1951).

\bibitem{still1} F. H. Stillinger, J. Chem. Phys. {\bf 65}, 3968 (1976).

\bibitem{still2} F. H. Stillinger and T. A. Weber, J. Chem. Phys. {\bf 68},
3837 (1978).

\bibitem{still3} F. H. Stillinger and T. A. Weber, Phys. Rev. B {\bf 22},
3790 (1980).

\bibitem{still4} F. H. Stillinger, J. Chem. Phys. {\bf 70}, 4067 (1979).

\bibitem{still5} F. H. Stillinger, Phys. Rev. B {\bf 20}, 299 (1979).

\bibitem{gerhard:00} S. Jorge, G. Kahl, E. Lomba, and J.~L.~F.~Abascal, 
J. Chem. Phys. {\bf 113}, 3302 (2000).

\bibitem{likos:prl:92} C. N. Likos and N. W. Ashcroft, Phys. Rev. Lett. {\bf 69},
316 (1992); Erratum, ibid., {\bf 69}, 3134 (1992).

\bibitem{likos:jcp:93} C. N. Likos and N. W. Ashcroft, J. Chem. Phys. {\bf 99},
9090 (1993).

\bibitem{wda1} W. A. Curtin and N. W. Ashcroft, Phys. Rev. A {\bf 32}, 2909 
(1985).

\bibitem{wda2} W. A. Curtin and N. W. Ashcroft, Phys. Rev. Lett. {\bf 56},
2775 (1986); Erratum, ibid., {\bf 57}, 1192 (1986).

\bibitem{mwda} A. R. Denton and N. W. Ashcroft, Phys. Rev. A {\bf 39}, 4701
(1989).

\bibitem{henderson} R. Evans, in {\it Fundamentals of Inhomogeneous Fluids},
ed. by D. Henderson (Marcel Dekker, New York 1992).

\bibitem{grewe1} N. Grewe and W. Klein, J. Math. Phys. {\bf 18}, 1729 (1977).

\bibitem{grewe2} N. Grewe and W. Klein, J. Math. Phys. {\bf 18}, 1735 (1977).

\bibitem{frisch:prl:85} H. L. Frisch, N. Rivier, and D. Wyler, 
Phys. Rev. Lett. {\bf 54}, 2061 (1985).

\bibitem{frisch:jcp:86} W. Klein and H. L. Frisch, J. Chem. Phys. {\bf 84},
968 (1986).

\bibitem{frisch:pra:87} H. L. Frisch and J. K. Percus, Phys. Rev. A {\bf 35},
4696 (1987).

\bibitem{frisch:pre:99} H. L. Frisch and J. K. Percus, Phys. Rev. E {\bf 60},
2942 (1999).

\bibitem{bagchi:jcp:88} B. Bagchi and S. A. Rice, J. Chem. Phys. {\bf 88},
1177 (1987).

\bibitem{ashcroft} N. W. Ashcroft and N. D. Mermin, {\it Solid State
Physics} (Holt Saunders, Philadelphia, 1976).

\bibitem{foot1} 
Strictly speaking, there is no freezing
{\it line} on the density-temperature plane but rather a domain
of coexistence, since it is equality of the grand potentials that
determines phase coexistence. Nevertheless, the condition of
equality of the Helmholtz free energies determines a line that
runs between the melting- and crystallization lines and therefore
serves as a reliable locator of the freezing transition.

\bibitem{lindemann} F. A. Lindemann, Phys. Z. {\bf 11}, 609 (1910).

\bibitem{shapiro} J. N. Shapiro, Phys. Rev. B {\bf 1}, 3982 (1970).

\bibitem{hv1} J.-P. Hansen and L. Verlet, Phys. Rev. {\bf 184}, 151 (1969).

\bibitem{hv2} J.-P. Hansen and D. Schiff, Mol. Phys. {\bf 25}, 1281 (1973).

\bibitem{dieter:jcp:05} D. Gottwald, G. Kahl, and C. N. Likos,
J. Chem. Phys. {\bf 122}, 204503 (2005).

\bibitem{bianca:long} B. M. Mladek, D. Gottwald, G. Kahl, M. Neumann,
and C. N. Likos, in preparation (2007).

\bibitem{foot2}
A very important exception
is the hard-sphere potential, for which no harmonic expansion
can be made. The fact that hard-sphere crystals do still
have density profiles that are very well modeled by Gaussians,
calls for a different explanation there. It can be argued that
the huge number of uncorrelated collisions with the neighbors,
together with the central limit theorem, are responsible for
the Gaussianity of density profiles in hard-sphere solids.

\bibitem{weeks:prb:81} J.~D.\ Weeks, Phys.\ Rev.\ B {\bf 24}, 1530 (1981).

\bibitem{lieb} E. Lieb, Rev. Mod. Phys. {\bf 48}, 553 (1976).

\bibitem{rex:molphys:06} M. Rex, C. N. Likos, H. L{\"o}wen, and 
J. Dzubiella, Mol. Phys. {\bf 104}, 527 (2006) and references therein.

\bibitem{bianca:dendris} B. M. Mladek, G. Kahl, and C. N. Likos,
in preparation (2007).

\end{thebibliography}
\end{document}